\documentclass[twocolumn,epjb,graphics,preprintnumbers,superscriptaddress,amsmath,amssymb]{revtex4}
\usepackage{graphicx}% Include figure files
\usepackage{dcolumn}% Align table columns on decimal point
\usepackage{bm}% bold math
\usepackage{amsmath,amssymb}
\newcommand{\be}{\begin{eqnarray}}
\newcommand{\ee}{\end{eqnarray}}

\usepackage[colorlinks,citecolor=blue]{hyperref}

\begin{document}

\title{Topological invariants, zero mode edge states and finite size effect for a generalized non-reciprocal Su-Schrieffer-Heeger model}
%\title{Phase diagram, zero mode edge states and finite size effect for a generalized non-reciprocal Su-Schrieffer-Heeger model}
%\title{Phase diagram, fate of zero mode edge states and finite size effect in a generalized non-reciprocal Su-Schrieffer-Heeger model}
%\title{Topological phases of a generalized non-reciprocal Su-Schrieffer-Heeger model: phase diagrams, finite size effect and fate of zero mode edge states}

\author{Hui Jiang}
\affiliation{Beijing National Laboratory for Condensed Matter Physics, Institute of Physics, Chinese Academy of Sciences, Beijing 100190, China}
\affiliation{School of Physical Sciences, University of Chinese Academy of Sciences, Beijing 100049, China}
%\author{Collaborators}
%\affiliation{***********, China}
\author{Rong L\"{u}}
\affiliation{Department of Physics, Tsinghua University, Beijing 100084, China}
\affiliation{Collaborative Innovation Center of Quantum Matter, Beijing, China}
\author{Shu Chen}
\email{schen@iphy.ac.cn}
\affiliation{Beijing National Laboratory for Condensed Matter Physics, Institute of Physics, Chinese Academy of Sciences, Beijing 100190, China}
\affiliation{School of Physical Sciences, University of Chinese Academy of Sciences, Beijing 100049, China}
\affiliation{Yangtze River Delta Physics Research Center, Liyang, Jiangsu 213300, China}
\date{\today}
\begin{abstract}
Intriguing issues  in one-dimensional non-reciprocal topological systems include the breakdown of usual bulk-edge correspondence and the occurrence of half-integer topological invariants.  In order to understand these unusual topological properties,  we investigate the topological phase diagrams and the zero-mode edge states of a generalized non-reciprocal Su-Schrieffer-Heeger model with a general form fulfilling the chiral symmetry, based on some analytical results. Meanwhile, we provide a concise geometrical interpretation of the bulk topological invariants in terms of two independent winding numbers and also give an alternative interpretation related to the linking properties of curves in three-dimensional space.  For the system under the open boundary condition, we construct analytically the wavefunctions of zero-mode edge states by properly considering a hidden symmetry of the system and the normalization condition with the use of biorthogonal eigenvectors. Our analytical results directly give the phase boundary for the existence of zero-mode edge states and unveil clearly the evolution behavior of edge states. In comparison with results via exact diagonalization of finite-size systems, we find our analytical results agree with the numerical results very well.
\end{abstract}

\maketitle

\section{introduction}
A characteristic feature of topological systems is the existence of robust edge states immune to symmetry-preserving perturbations \cite{Kane2010,Qi,Kitaev2009}.
In general, the emergence of edge states under the open boundary condition (OBC) is attributed to the existence of a nontrivial topological invariant in the bulk system,
which is referred to as the bulk-boundary correspondence \cite{Kane2010,Qi}. Recently much attention has been drawn to non-Hermitian topological systems \cite{Rudner,Hu2011,Esaki2011,Yin2018,Zhu2014,Gong2018, Liu2019,Lee2016,Leykam2017,Xiong2018,Yao2018,Yao20181,Kunst2018,Shen2018,Liang2013,Lieu2018,Jiang2018,Das}, which can be viewed as a direct generalization of topological band systems by releasing the Hermitian constraint \cite{Gong2018,Sato,LiuCH,Zhou}. Non-Hermitian systems have been found to be good candidates for describing some open non-equilibrium quantum systems \cite{Carmichael1993,Rotter2009,Verstraete2009,Harari2018}, photonic and acoustic systems with gain and loss \cite{Diehl2011,Feng2014,Hodaei2014,Gao2015,Xu2016,Chen2017,Charles2017,Ozawa2019} and electronic circuits \cite{Wang2018,Jiang2019,Ezawa2019,Ezawa20191}.
It has been demonstrated that the non-Hermitian systems display some peculiar properties without Hermitian correspondence, e.g., complex eigenvalues, biorthogonal eigenvectors and the existence of exception points, etc \cite{Hodal2017,Alvarez2018,Zhu2014,Yuce2015,Yuce2016,Menke2017,Xu2017,Cerjan2018,Zyuzin2018,Lieu20181,Bergholtz2018,Zhou2018,wang2019,Yoshida2019,Budich2019,Moors2019,Okugawa2019,Yang2019}.
For topological non-Hermitian systems, recent studies have unveiled that the bulk-boundary correspondence does not always hold true \cite{Alvarez2018,Xiong2018,Yao2018}, the unusual bulk-boundary correspondence and non-Hermitian skin effect may emerge in some non-reciprocal systems \cite{Yao2018,Jiang2019}.\\
\indent Despite its simplicity, the Su-Schrieffer-Heeger (SSH) model \cite{Schrieffer1979} and its extensions \cite{Rice1982,linhu2014,Zvyagin} have attracted extensive studies in the past decades as it can be used as a playground for illustrating rich topological phenomena. Recently, a non-Hermitian SSH model with chiral symmetry was proposed \cite{Lieu2018,Yin2018} and it is shown that this model displays rich phase diagrams with phases characterized by half-integer topological numbers \cite{Yin2018}. While a geometrical interpretation of bulk topological invariant is given in Ref.\cite{Yin2018,Jiang2018}, further studies of the model under the OBC unveil the existence of non-Hermitian skin states and breakdown of the conventional bulk-boundary correspondence \cite{Yao2018}, i.e., the region characterized by the nontrivial bulk topological invariant is different from the region for the existence of zero-mode edge states. The unusual bulk-boundary correspondence has stimulated intensive studies on the underlying physical meanings and reasons  \cite{Yuce2019,Jin2019,Lee2019,Herviou2019,Zirnstein2019,Pocock2019,Borgnia2019,Yokomizo2019,ZhaoYX2019,Lee2018,Edvardsson2019,Jiang2019,Song2019,Kunst2019,WuHC2019}.
Recently, there are also some interesting experimental studies associated with extended non-Hermitian SSH models \cite{Takata,arXiv:1907.11619,XuePeng}. While topological insulating phase is experimentally observed in non-Hermitian optical lattices with  parity-time symmetry induced solely by gain and loss control \cite{Takata}, non-Hermitian version of the SSH chain with chiral symmetry has been realized in robotic metamaterials \cite{arXiv:1907.11619}, and non-Hermitian bulk-boundary correspondence in discrete-time non-unitary quantum-walk dynamics of single photons has also been observed \cite{XuePeng}.\\
\indent In order to understand the breakdown of conventional bulk-boundary correspondence in non-Hermitian topological systems, one of the key issues is the understanding of the fate of zero mode edge states in the presence of non-reciprocal hopping processes, which may induce the non-Hermitian skin effect. By considering the semi-infinite boundary condition, it was demonstrated that the existence of left and right zero-mode edge states is
consistent with the bulk topological numbers \cite{Yin2018}, which however is contradicted to numerical results via the diagonalization of finite-size systems \cite{Yao2018,Kunst2018}. This contradiction suggests that the semi-infinite zero mode solutions no longer hold true for the finite-size system. When a finite-size chain is considered, the left and right zero-mode edge states are coupled together accompanying with the opening of a finite gap \cite{ZhouBin}. Although the numerical results have unveiled the discrepancy between non-Hermitian zero-mode states and the conventional zero-mode states, it is still puzzling to understand why the semi-infinite solutions fails to match the numerical solution of finite-size system even in the large size limit, as it should be in the conventional Hermitian counterpart.
Furthermore, the skin effect suggests the zero-mode states are either on the left or the right boundary,  and thus it is still a puzzling problem for understanding the transition process from a unified zero-mode solution.\\
\indent To get a deep understanding for the fate of zero-mode edge states in a finite non-reciprocal topological system, it is highly desirable to explore an analytical form of zero-mode states which can help us clarify the puzzling problems and give quantitative predication of wavefuntions and finite-size gap which is consistent with the numerical results. To this end, in this work we study a generalized non-reciprocal Su-Schrieffer-Heeger model and give an analytical form of ansatz wavefuntions, which are taken as the superposition of semi-definite zero-mode solutions. By analyzing the symmetry of the system, we find that the existence of a hidden symmetry plays an important role in fixing the form of superposition coefficients. Taking account of the hidden symmetry  and normalization condition by using biorthogonal eigenvectors, our zero-mode wavefunctions are uniquely determined without any variational parameter. In comparison with results via numerical diagonalization of finite-size systems, we show that our analytical results agree with the numerical results very well.\\
\indent The paper is organized as follows. In Sec. II, we introduce the generalized non-reciprocal Su-Schrieffer-Heeger model and determine its phase diagram via the calculation of topological invariant. The general geometrical meaning of topological invariants in momentum $k$ space is also discussed. In Sec.III, we focus  on the study of the zero mode edge state under OBC. We give analytically the condition for the occurrence of zero mode edge states,
and give explicitly the analytical expression of zero-mode edge states at the finite-size system by enforcing the wavefunctions to fulfill the hidden symmetry of the system.
A summary is given in Sec. IV.\\
\section{Model, topological invariant and phase diagram}
\begin{figure}[bp]\label{angel}
	\centering
	\includegraphics[width=0.8\columnwidth]{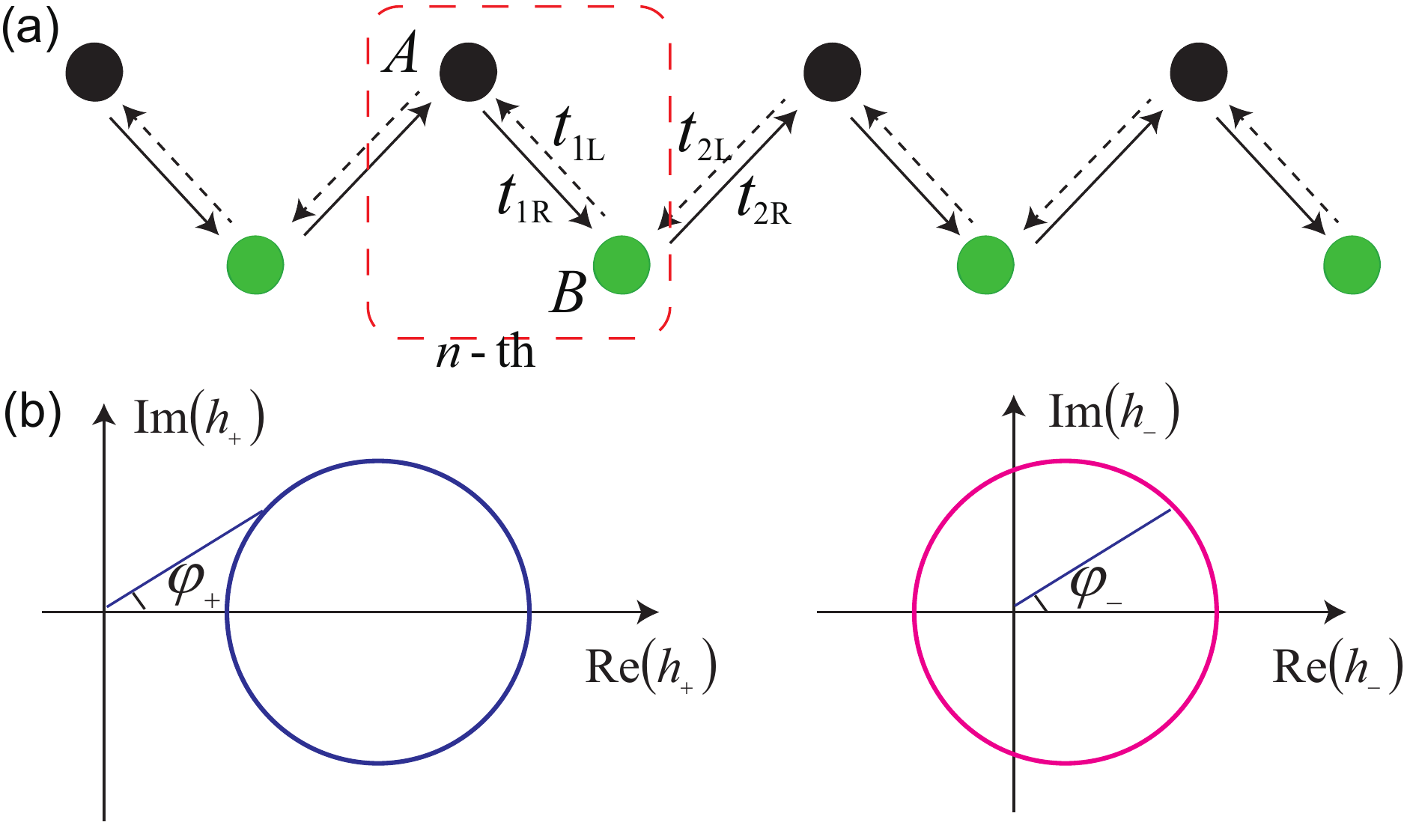}
\caption{(a) Schematic diagram of the generalized non-reciprocal SSH model. (b) Geometric configurations of topological invariants $\nu_{\pm}$ corresponding to $\nu_{+}=0$ and $\nu_{-}=1$. The trajectory of $h_{\pm}(k)$ forms a close curve either encircling or not encircling around the origin point when the momentum $k$ goes across the Brillouin region.} \label{angle}
\end{figure}

Consider a general non-reciprocal one-dimensional (1D) non-Hermitian model described by
\begin{equation}\label{non-ssh}
\begin{split}
  H=\sum_{n} t_{{1\text{L}}} &|n,A\rangle\langle n,B|+t_{{2\text{R}}} |n, A\rangle\langle n-1, B|\\
  &+t_{{1\text{R}}}|n,B\rangle\langle n,A|+t_{{2\text{L}}} |n,B\rangle\langle n+1,A|,
  \end{split}
\end{equation}
where $t_{{1(2) \text{R(L)}}}$ is the right (left) intra (inter)-hopping amplitude as shown schematically in Fig.\ref{angle}(a), $A~(B)$ represents the sublattice labels and $n$ indicates the $n$-th cell of the lattice. For the system under the periodic boundary condition (PBC), it is convenient to get the Hamiltonian in the momentum space, which can be represented as
\begin{equation}\label{model}
H(k)= \sum_k   \psi_k^{\dagger} h(k) \psi_k,
\end{equation}
where $\psi_k=(\langle k,A|,\langle k,B|)^{T}$, and \begin{equation}\label{chiral}
  h(k)=\left(
              \begin{array}{cc}
                0 & h_+(k) \\
                h_-(k) & 0 \\
              \end{array}
            \right),
\end{equation}
with
\begin{equation} \label{hk-SSH}
  \begin{split}
  h_+(k)&=t_{{1\text{L}}}+t_{2\text{R}}e^{-ik},\\
   h_-(k)&=t_{1\text{R}}+t_{2\text{L}} e^{ik}.
  \end{split}
\end{equation}
Eq.(\ref{chiral}) can be alternatively written as
\begin{equation}
h(k)= h_x(k) \sigma_x + h_y(k) \sigma_y,
\end{equation}
with $h_+ =h_x -i h_y$ and $h_-=h_x+i h_y$.
It is obvious that the Hamiltonian satisfies the chiral symmetry
\[
Uh(k)U^{\dagger} =-h(k),
 \]
with unitary matrix $U=\sigma_z$. The model Eq.(\ref{non-ssh}) can be viewed as a generalized non-reciprocal SSH model with the most general form.  When $t_{2\text{L}}=t_{2\text{R}}$, the model reduces to the non-Hermitian SSH model studied in previous references \cite{Yin2018,Yao2018}. If both $t_{1\text{L}}=t_{1\text{R}}$ and $t_{2\text{L}}=t_{2\text{R}}$ are fulfilled, the model reduces to the standard SSH model \cite{Schrieffer1979}.

\indent It is straightforward to see that the eigenvalues $E_{1,2}$ of $h(k)$ fulfill
\begin{equation}\label{eigenvalue}
E^2_{1,2}(k) =h_+({k})\cdot h_-({k}),
\end{equation}
where the eigenvalue satisfies $E_1=-E_2$.  The eigenvalues are generally complex and the corresponding eigenvectors are given by
 $|\psi_{1,2}\rangle=1/\sqrt{2}\left(
 \begin{array}{cc}
  h_{+}/E_1, & \pm1 \\
 \end{array}
 \right
 )^{T}$ where $+1$ and $-1$ corresponding to $|\psi_{1}\rangle$ and $|\psi_{2}\rangle$, respectively. It's easy to find that the relevant left vector $\langle\phi_{1,2}|$, which fulfills  $\langle\phi_{1,2}| h(k)=E_{1,2}\langle\phi_{1,2}| $ (or $ h^{\dagger}(k)|\phi_{1,2} \rangle= E^*_{1,2} |\phi_{1,2} \rangle  $ ), is given by
 $\langle\phi_{1,2}|=1/\sqrt{2}\left(
 \begin{array}{cc}
  h_{-}/E_1, & \pm1 \\
 \end{array}
 \right
 )$.  The biorthogonal eigenvectors fulfill $\langle\phi_{i}|\psi_{j}\rangle=\delta_{i,j}$ with $i,j=1,2$. The topological invariance related to Berry phase is \[
 \nu_{s,j}=\frac{1}{\pi} \int \text{d} {\bm{k}} \langle\phi_{j}|i\partial_{\bm{k}}|\psi _{j}\rangle,
 \]
 where the  subscript  $j=1,2$ represents the band index. It is easy to check $\nu_{s,1}=\nu_{s,2}$, thus we can omit the band index. After some algebras, we can represent $\nu_{s}$ as
 \begin{equation}\label{Berry pahse}
   \nu_s= \frac{1}{2}(\nu_- - \nu_+),
 \end{equation}
where
\begin{equation}
\nu_{\pm}=\frac{1}{2\pi} \oint \partial_{k} \varphi_{\pm} \text{d} {k},  \label{nu+-}
\end{equation}
and the angles $\varphi_{\pm}$ is defined by  $h_{\pm}=|h_{\pm}|e^{i\varphi_{\pm}}$ or alternatively by
\begin{equation}
\tan  \varphi_{\pm} = \frac{\text{Im}(h_{\pm}(k))}{\text{Re}(h_{\pm}(k))},
\end{equation}
as schematically displayed in Fig.\ref{angle}(b).
In terms of $\nu_{\pm}$, it is straightforward that the winding of energy
\[
\nu_E=\frac{1}{2\pi}\int \text{d} k \partial_k \text{Arg} (E_{1,2})
\]
can be also be represented as
\begin{equation}
  \nu_E=\frac{1}{2}(\nu_{+}+\nu_{-}).
\end{equation}

Next we discuss the geometrical interpretation of the topological invariants.
From Eq.(\ref{eigenvalue}), we can see that  $h_{+}=0$ and $h_{-}=0$ correspond to two exception points of the chiral non-Hermitian system. When $k$ goes cross the  Brillouin region, the trajectory of $h_{\pm}(k)$ projected in the two-dimensional space spanned by Re($h_{\pm}(k)$) and Im($h_{\pm}(k)$) forms a closed curve, either encircling or not around the origin as shown like Fig.1(b).
%And EPs can be represented the origin of the coordinate (Re$(h_{\pm})$, Im$(h_{\pm})$)system,
According to the definition of Eq.(\ref{nu+-}), $\nu_{+}$ and $\nu_{-}$ denote the winding numbers of the closed curves  encircling the
exceptional points $h_{+}=0$ and $h_{-}=0$, respectively.
\begin{figure}[bp]
	\centering
	\includegraphics[width=1\columnwidth]{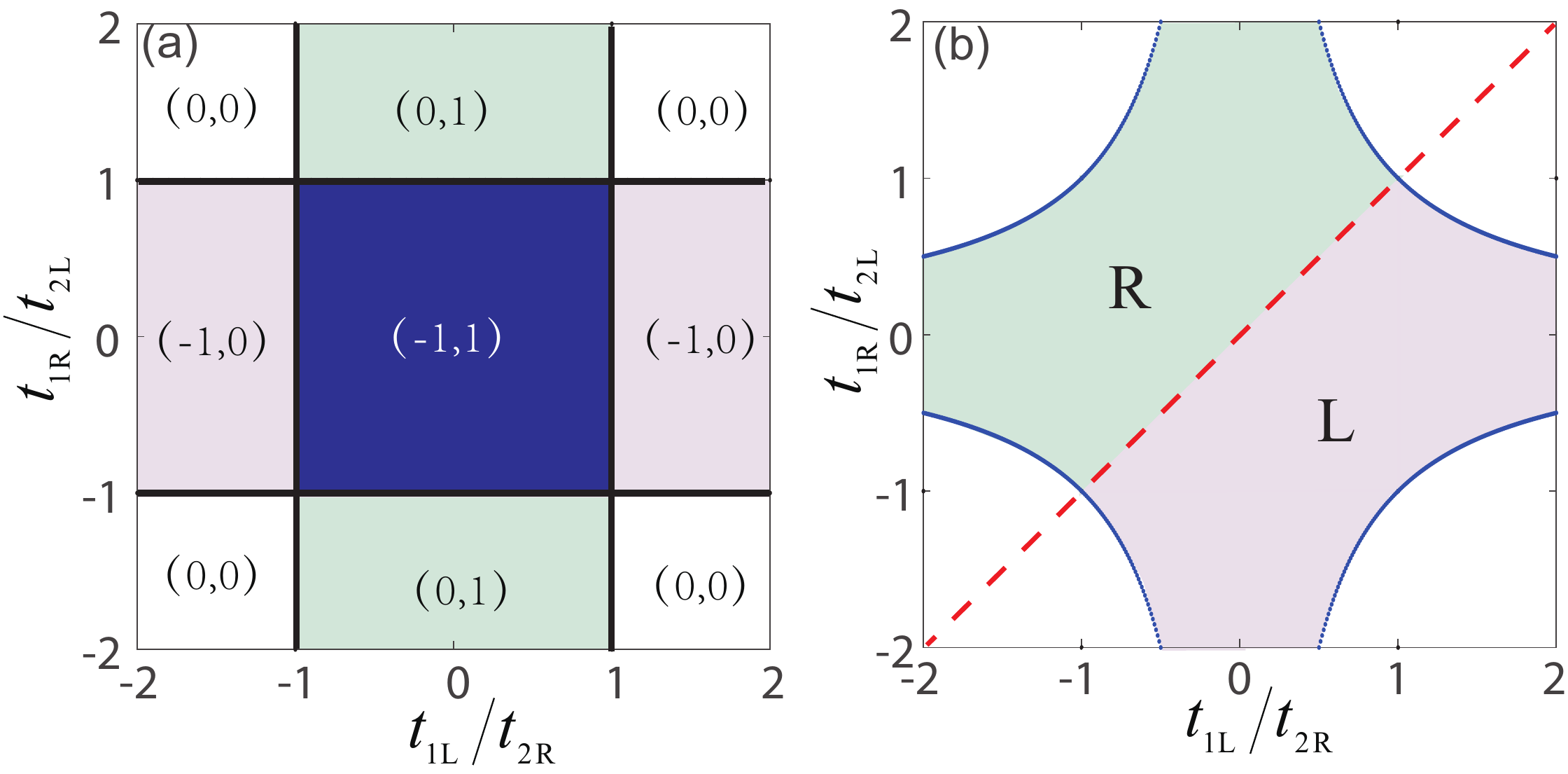}
\caption{Phase diagram for the system under the PBC (a) or OBC (b). (a) Topologically different phases are characterized  by topological invariants ($\nu_+,\nu_-$). Phase boundaries are denoted by the black lines, which are also the band touching lines. (b) Topological or trivial phase is characterized by the existence or absence of zero-mode edge states with the phase boundaries denoted by the blue curves. The red dashed lines distinguish the skin effect of bulk states, which are located at the left or right edge in the area below or above the line.  }\label{phase}
\end{figure}

For the generalized non-reciprocal SSH model, we fix parameters $t_{1(2)\text{R(L)}}$  to be real without loss of generality, after some straightforward calculations (see the detail in Appendix.\ref{a}), the topological invariant $\nu_{\pm}$ can be given by
\begin{equation}\label{nu}
  \begin{split}
  \nu_+&=\frac{1}{2}\{\mathrm{sgn}(t_{1\text{L}}- t_{2\text{R}})- \mathrm{sgn}(t_{1\text{L}}+t_{2\text{R}})\},\\
   \nu_-&=\frac{1}{2}\{\mathrm{sgn}(t_{1\text{R}}+ t_{2\text{L}})- \mathrm{sgn}(t_{1\text{R}}- t_{2\text{L}})\},
  \end{split}
\end{equation}
where the sign function $\mathrm{sgn}(x)=1$ for $x>0$ and $\mathrm{sgn}(x)=-1$ for $x<0$.
In Fig.\ref{phase}(a), we show the phase diagram of the model (Eq.(\ref{non-ssh})) with different phases characterized by different $\nu_{\pm}$.
The phase diagram is plotted in the parameter space spanned by $t_{1\text{R}}/ t_{2\text{L}}$ and $t_{1\text{L}}/t_{2\text{R}}$, and the different topological invariant $(\nu_{+},\nu_{-})$ is marked on Fig.\ref{phase}(a). Due to the existence of chiral symmetry, the phase boundaries ($|t_{1\text{R}}/ t_{2\text{L}}|=|t_{1\text{L}}/t_{2\text{R}}|=1$) correspond to the band touching points determined by $E_{1,2}=0$.

% only if $t_{1\text{L}}=\pm t_{2\text{R}}$ ($h_+(k_{\text{c}})=0$) or $t_{1\text{R}}=\pm t_{2\text{L}}$ ($h_-(k_{\text{c}})=0$), the eigenvalues $E_{1,2}$ are equal to each other with the momentum $k_{\text{c}}$($k_{\text{c}}=0,\pi$), namely, the non-Hermitian system has EPs.

%\indent In addition, the condition for the existence of EPs can be alternatively written as $\text{Re}(h_x)=\mp\text{Im}(h_y)$ and $\text{Re}(h_y)=\pm\text{Im}(h_x)$, which is a deformation of function $h_{\pm}=0$.In the coordinate system of (Re$(h_{x})$, Re$(h_{y})$), the curves ($\mp\text{Im}(h_y),\pm\text{Im}(h_x)$) with evolution of  momentum $k$ form exception rings, corresponding to the two exception points, respectively. As shown in Fig.{\ref{yin1}},  the purple and red curve represent $h_{\pm}=0$, respectively.

\section{zero mode edge states}\label{edge}
Considering the system under the open boundary condition, it is convenient to represent the state $|n,A \rangle$ and $|n,B \rangle$ as product of vectors defined in the space of position $n=(1,2,..L)$ and sublattice $A$ and $B$, i.e., $|n\rangle\otimes|\xi\rangle$ with $|\xi_a\rangle = (1, 0)^T$ and  $|\xi_b\rangle = (0, 1)^T$ corresponding to the $A$ and $B$ sublattice. In terms of these terminologies \cite{ZhaoYX2019}, the Hamiltonian under the OBC can be rewritten as
\begin{equation}\label{OBC1}
   \small \tilde{\mathcal{H}}=\text{I}\otimes\left(
                             \begin{array}{cc}
                               0 &t_{1\text{L}} \\
                               t_{1\text{R}} & 0 \\
                             \end{array}
                           \right)
                           +\hat{S}\otimes\left(
                             \begin{array}{cc}
                               0 & t_{2\text{R}}\\
                               0 & 0 \\
                             \end{array}
                           \right)
                           +\hat{S}^{\dagger}\otimes\left(
                             \begin{array}{cc}
                               0 & 0 \\
                              t_{2\text{L}} & 0 \\
                             \end{array}
                           \right),
  \end{equation}
  with  unit operator I, backward and forward translation
operators defined by $\hat{S}|i\rangle=|i+1\rangle$ and $\hat{S}^{\dagger}|i\rangle=|i-1\rangle$, respectively. Explicitly, we have $\hat{S}= \sum_n |n+1\rangle\langle n|$ and $\hat{S}^{\dagger}=\sum_n |n-1\rangle\langle n|$.

%As $S^{\dagger}_{i,j}=\langle i|\hat{S}|j\rangle=\delta_{i,j+1}$ and $S^{\dagger}_{i,j}=\langle i|\hat{S}^{\dagger}|j\rangle=\delta_{i,j-1}$.

If we consider the semi-infinite limit from the left boundary, we can get the zero-mode eigenstate of the form:
\begin{equation} \label{psia}
  |{\psi}_a\rangle=1/\mathcal{N}_a\sum_{n=1}^{L-1}\beta_a^{n-1}|n\rangle\otimes|\xi_a\rangle,
\end{equation}
with $|\xi_a\rangle=(1,0)^\text{T}$ and $\beta_a=-t_{1\text{R}}/t_{2\text{L}}$ (see Appendix B).
Similarly, the zero-mode eigenstate of $\tilde{\mathcal{H}}^{\dagger}$ is given by
\begin{equation*}
  |{\phi}_a\rangle=1/\mathcal{N}_a\sum_{n=1}^{L-1}{\beta'}_a^{n-1}|n\rangle\otimes|\xi_a\rangle,
\end{equation*}
with
$\beta'_a=-t_{1\text{L}}/t_{2\text{R}}$. The normalization constant is determined by
$\langle \phi_a |\psi_a \rangle=1$, which gives rise to $\mathcal{N}_a=\sqrt{(1-(\beta'_a\beta_a)^L)/(1-\beta'_a\beta_a)}$. From Eq.(\ref{psia}), we see that the state is an edge state exponentially decaying from  the left boundary as long as $|\beta_a|<1$, i.e.,
\begin{equation}
\left| \frac{t_{1\text{R}}}{t_{2\text{L}}} \right|<1. \label{LZM1}
\end{equation}
The requirement that the normalization constant should be a finite  number gives an additional constraint condition $|\beta_a \beta'_a| < 1$, i.e.,
\begin{equation}
|t_{1\text{L}}t_{1\text{R}}|<|t_{2\text{L}}t_{2\text{R}}|. \label{LZM2}
\end{equation}
Only when both Eq.(\ref{LZM1}) and (\ref{LZM2}) are fulfilled, the left zero mode solution exits.

In the same way,  considering the semi-infinite limit from the right boundary, we have the right zero-mode state of the following form:
\begin{equation} \label{psib}
  |{\psi}_b\rangle=1/\mathcal{N}_b\sum_{n=0}^{L-1}\beta_b^{n}|L-n\rangle\otimes|\xi_b\rangle,
\end{equation}
with $|\xi_b\rangle=(0,1)^\text{T}$ and $\beta_b=- t_{1\text{L}}/t_{2\text{R}}=\beta'_a$.
Similarly, we have the zero-mode eigenstate of $H^{\dagger}$
\begin{equation*}
  |{\phi}_b\rangle=1/\mathcal{N}_b\sum_{n=0}^{L-1}{\beta'_b}^{n}|L-n\rangle\otimes|\xi_b\rangle,
\end{equation*}
where $\beta'_b= -t_{1\text{R}}/t_{2\text{L}} = \beta_a$ and the normalization constant
$\mathcal{N}_b=\sqrt{(1-(\beta'_b\beta_b)^L)/(1-\beta'_b\beta_b)}=\mathcal{N}_a$ is determined by $\langle{\phi}|{\psi}\rangle=1$.
Eq.(\ref{psib}) suggests that the state is an edge state exponentially decaying from  the right boundary as long as $|\beta_b|<1$, i.e.,
\begin{equation}
\left| \frac{t_{1\text{L}}}{t_{2\text{R}}} \right|<1. \label{RZM1}
\end{equation}
The requirement that the normalization constant should be a finite  number gives an additional constraint condition $|\beta_b \beta'_b| < 1$, which is identical to Eq.(\ref{LZM2}) due to $|\beta_b \beta'_b|=|\beta_a \beta'_a|$.
%\begin{equation}
%|t_{1\text{L}}t_{1\text{R}}| < |t_{2\text{L}}t_{2\text{R}}|, \label{RZM2}
%\end{equation}
Only when both Eq.(\ref{RZM1}) and (\ref{LZM2}) are fulfilled, the right zero mode solution exits.

\begin{figure}[tbp]
	\centering
	\includegraphics[width=1\columnwidth]{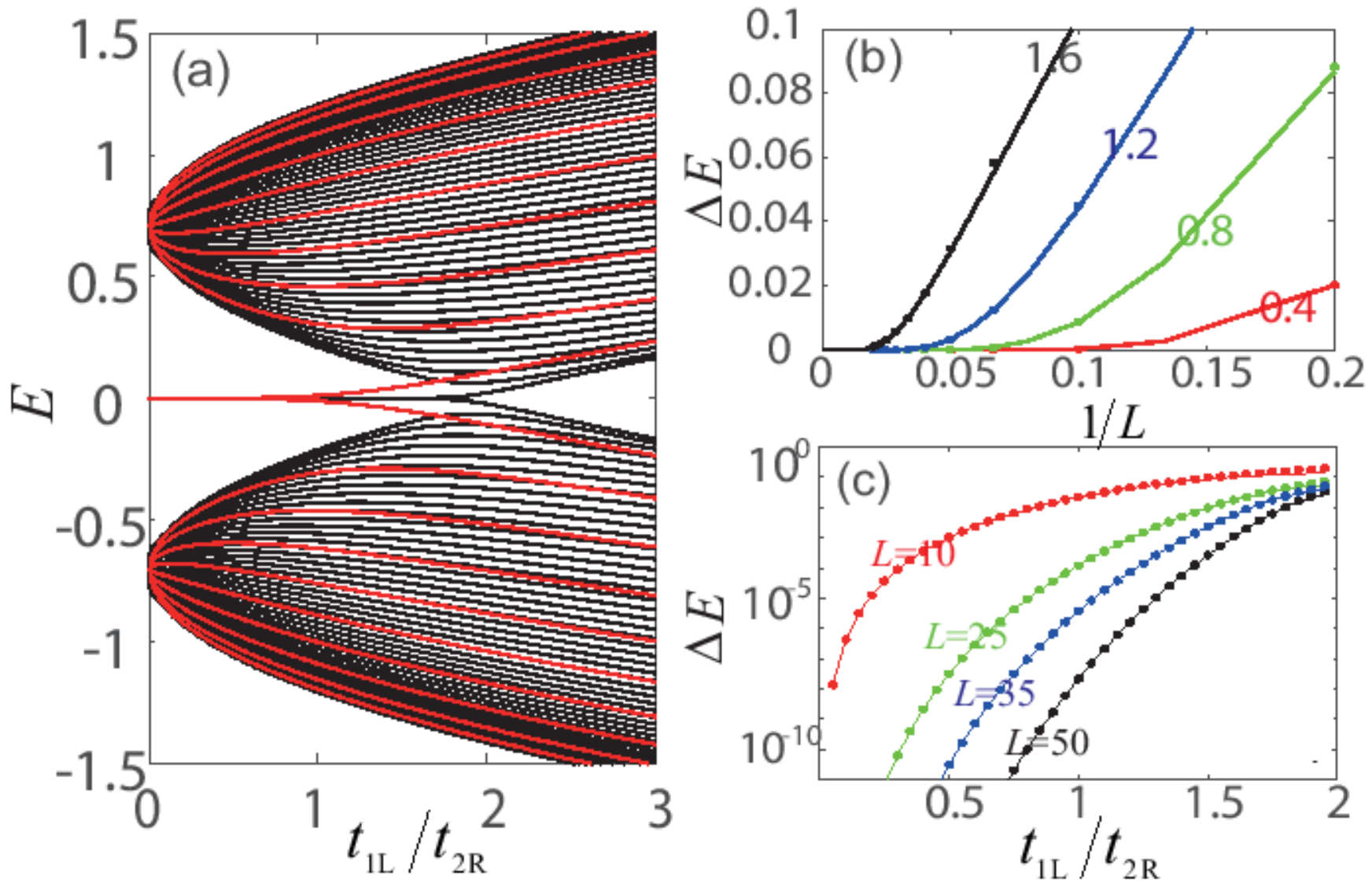}
\caption{(a) The spectra of the generalized non-reciprocal SSH
model under OBC.  The red curves represent the system with the length $L=10$  and the black ones with $L=50$; (b) The finite-size
gap $\Delta E$ versus $1/L$ for various $t_{1\text{L}}/t_{2\text{R}}=0.4$, $0.8$, $1.2$ and $1.6$. (c) $\Delta E$ versus $t_{1\text{L}}/t_{2\text{R}}$ for various $L$. While dots represent the numerical results,  lines denote results obtained analytically. Here we fix $t_{2\text{R}}=1$ as the unit of energy and take $t_{2\text{L}}=0.5$ and $t_{1\text{R}}/t_{2\text{L}}=0.5$.}\label{1E}
\end{figure}

From the above discussion, it is known that regions for the existence of zero mode edge states are determined by
\begin{equation}
\left| \frac{t_{1\text{R}}}{t_{2\text{L}}}\times \frac{t_{1\text{L}}}{t_{2\text{R}}}  \right|< 1. \label{phaseboundary}
\end{equation}
By using the above equation, we can determine the phase boundaries of the system under OBC, as shown in Fig.\ref{phase}(b).
When $t_{2\text{L}}=t_{2\text{R}}$, our results are consistent with the boundary conditions obtained by Kunst et. al \cite{Kunst2018}.
The  zero mode states $|{\psi}_{a(b)}\rangle$ distributes only at the A (or B) sublattice and is the eigenstate of the system only when $L \rightarrow \infty$.  For a finite-size system, these zero mode states are no longer eigenstates of the system. In general, the left and right edge states couple together and open a tiny gap due to the finite-size effect.\\
\indent To see the finite size effect, we numerically diagonalize the Hamiltonian with different sizes of $L$. As shown in Fig.\ref{1E} (a), the degeneracy of zero modes is lifted accompanying with the opening of a finite gap. To get an intuitive understanding of the fate of zero modes, we take the wave-function of the finite size system as the superposition of left and right edge states, i.e.,
\begin{equation}
|\psi \rangle = |\psi_{a}\rangle + c |\psi_{b}\rangle, \label{trial}
\end{equation}
where $c$ is a constant which can be determined by considering the symmetry of the system.
To see it clearly, we notice that there exists a hidden symmetry for the non-Hermitian Hamiltonian. We find that  the operator $P$, defined by
  \begin{equation}
    P=\sum_{n=1}^{L} r^{L-2n+1}|L-n+1\rangle\langle n|\otimes \left(
                                                               \begin{array}{cc}
                                                                 0 & \alpha^{-1} \\
                                                                  \alpha & 0 \\
                                                               \end{array}
                                                             \right),
  \end{equation}
  with $r=\sqrt{t_{1\text{R}}t_{2\text{R}}/t_{1\text{L}}t_{2\text{L}}}$ and $\alpha=\sqrt{t_{1\text{R}}/t_{1\text{L}}}$, commutates with $\tilde{\mathcal{H}}$, i.e.,
  \[
  [P,\tilde{\mathcal{H}}]=0.
  \]
  In other words,  the non-degenerate eigenvector of the non-Hermitian Hamiltonian $ \tilde{\mathcal{H}}$ should be simultaneously the eigenvector of the operator $P$.  Therefore, we require the ansatz wave functions to be the eigenfuntion of the operator $P$, which fixes the parameter $c$ in Eq.(\ref{trial}) and leads to
    \begin{equation}\label{phi}
   \begin{split}
     |\psi_{1}\rangle &=\frac{1}{\sqrt{\langle\phi_{1}|\psi_{1}\rangle}}( |\psi_{a}\rangle+ r^{{L-1}}\cdot\alpha|\psi_{b}\rangle) ,\\
     |\psi_{2}\rangle &= \frac{1}{\sqrt{\langle\phi_{2}|\psi_{2}\rangle}}(|\psi_{a}\rangle-r^{{L-1}}\cdot\alpha|\psi_{b}\rangle) ,\\
     \langle\phi_{1}|&=\frac{1}{\sqrt{\langle\phi_{1}|\psi_{1}\rangle}}(r^{{L-1}}\cdot\alpha\langle\phi_{a}|+\langle\phi_{b}|) ,\\
     \langle\phi_{2}| &= \frac{1}{\sqrt{\langle\phi_{2}|\psi_{2}\rangle}}(r^{{L-1}}\cdot\alpha \langle\phi_{a}|-\langle\phi_{b}|),
     \end{split}
   \end{equation}
  with $\langle{\phi}_{i}|{\psi}_{i'}\rangle=\delta_{ii'}$ $(i,i'=1,2)$  and   $\langle{\phi}_{i}|H|{\psi}_{i'}\rangle=0$ for $i\neq i'$. It is easy to check that $P^2=I$ and $P|\psi_{1}\rangle=|\psi_{1}\rangle$ and $P|\psi_{2}\rangle=-|\psi_{2}\rangle$ (see detail in Appendix \ref{C11}). In the limit case with $t_{1,2}=t'_{1,2}$, our model reduces to the SSH model and the operator $P$ is reduced to an inverse operator.
 By using Eq.(\ref{phi}), it is straightforward to calculate
  \[
  E_{1,2}=\langle\phi_{1,2}|\tilde{\mathcal{H}}|\psi_{1,2}\rangle,
  \]
where we have $E_1=-E_2$ and the energy splitting is given by $\Delta E =E_1-E_2$. After some algebras, we can get
  \begin{equation}\label{deviation1}
\begin{split}
  \Delta E&=\frac{2\sqrt{t_{1\text{L}}t_{1\text{R}}}}{\mathcal{N}^2_a}\times\left (\sqrt{\frac{t_{1\text{R}}t_{1\text{L}}}{t_{2\text{L}}t_{2\text{R}}}}\right)^{L-1} ,\\
  \end{split}
\end{equation}and the
$1/\mathcal{N}^2_a=(1-(t_{1\text{L}}t_{1\text{R}}/t_{2\text{L}}t_{2\text{R}}))/(1-(t_{1\text{L}}t_{1\text{R}}/t_{2\text{L}}t_{2\text{R}})^L)$.
 And only when the system satisfies the condition $|t_{1\text{L}}t_{1\text{R}}|<|t_{2\text{L}}t_{2\text{R}}|$, the trial wave functions $|\psi_{1,2}\rangle$ makes sense. Meantime, the zero-energy deviation  $\Delta E$ decays exponentially with increasing length $L$, the decay rate is inversely proportional to $t_{1\text{L}}t_{2\text{L}}/t_{1\text{R}}t_{2\text{R}}$ as shown in Fig.\ref{1E}(b,c).
Comparing the gap sizes given by Eq.(\ref{deviation1}) with the numerical results obtained by exact diagonalization, we find that they agree very well as shown in Figs. \ref{1E}(b) and \ref{1E}(c).
\begin{figure}[bp]
	\centering
	\includegraphics[width=0.9\columnwidth]{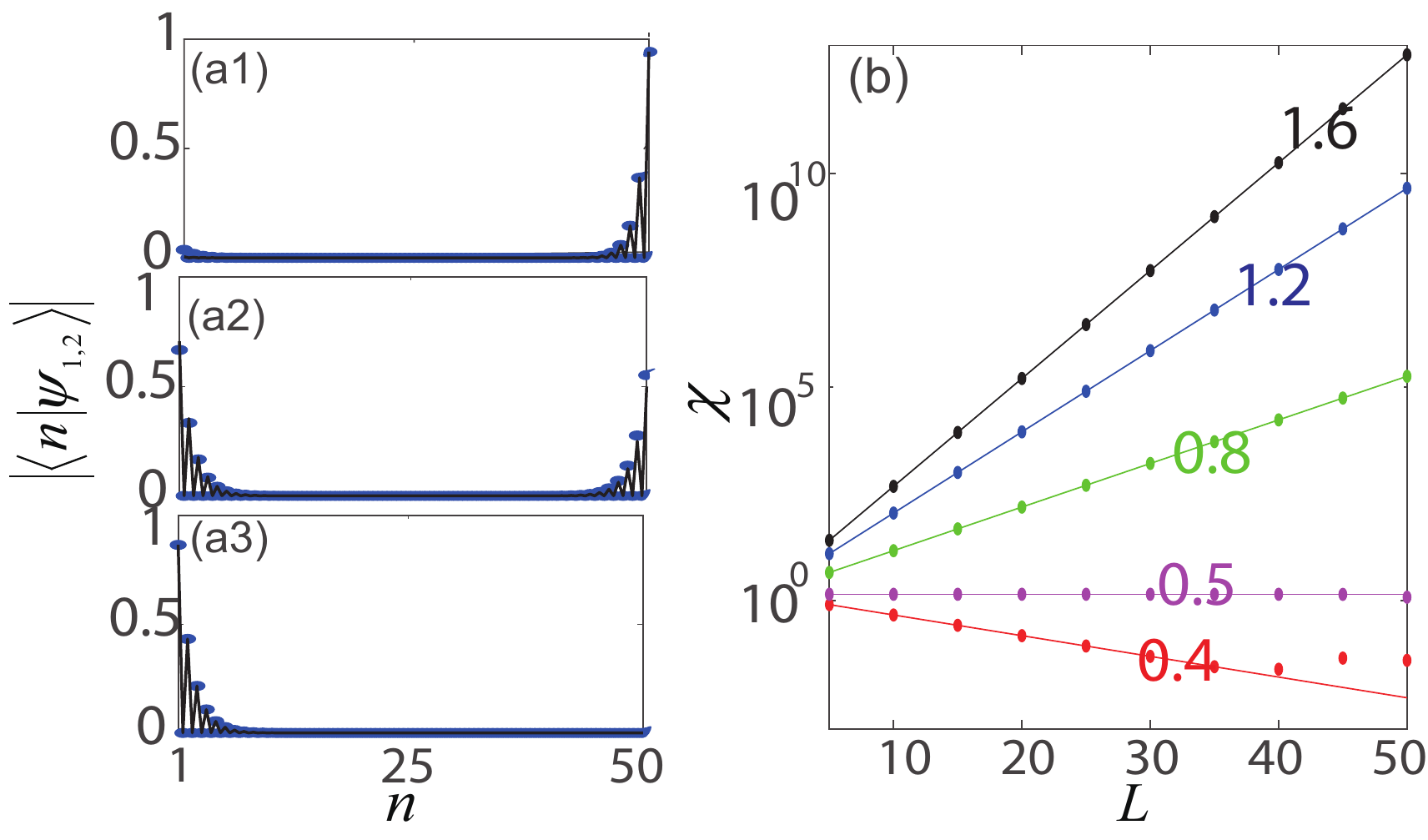}
\caption{(a1)-(a3) The distribution of zero mode states $|\psi_{1,2}\rangle$ with  $t_{1\text{L}}/t_{2\text{R}}=0.4$ (a1), $t_{1\text{L}}/t_{2\text{R}}=0.5$ (a2), $t_{1\text{L}}/t_{2\text{R}}=0.8$ (a3), obtained numerically (dots) and  analytically (lines); (b) The ratio $\chi$ versus $L$ with different $t_{1\text{L}}/t_{2\text{R}}$ obtained numerically (dots) and analytically (lines). The number marked in (b) represents the value of $t_{1\text{L}}/t_{2\text{R}}$  for the corresponding line. Here we fix $t_{2\text{R}}=1$ and take $t_{2\text{L}}=0.5$ and $t_{1\text{R}}/t_{2\text{L}}=0.5$. }\label{wave}
\end{figure}

In order to test the accuracy of ansatz wavefuntions, next we compare them with the numerical results. In Figs.\ref{wave}(a1)-(a3), we show the distributions of zero mode states $|\langle n |\psi_{1,2}\rangle|$ with $t_{1\text{L}}/t_{2\text{R}}=0.4$, $0.5$ and $0.8$, respectively, obtained by both analytical and numerical calculation. It is shown that they match very well.
 To describe the different probability distribution between at site $1,A$  and at site $L,B$, we define the ratio $\chi_{1,2}$ by
 \begin{equation*}
  \chi_{1,2}=\frac{|\langle 1,A|\psi_{1,2}\rangle|}{|\langle L,B|\psi_{1,2}\rangle|}.
 \end{equation*}
From the analytical forms of zero-mode wavefuntions, we see $\chi_{1}=\chi_2=\chi$. Numerically, we also find no difference for $\chi_{1}$ and $\chi_{2}$. By using Eq.(\ref{phi}), it then follows
\begin{equation}
\chi=\left(\frac{t_{1\text{L}}}{t_{1\text{R}}}\right)^{\frac{1}{2}} \left( \frac{t_{1\text{L}}t_{2\text{L}}}{t_{1\text{R}}t_{2\text{R}}} \right)^{\frac{L-1}{2}}. \label{chi}
\end{equation}
From the above expression, we see that the zero mode wavefunction would be located at the left or right edge  depending on $\left|{t_{1\text{L}}} / {t_{2\text{R}}} \right| > \left| {t_{1\text{R}}}/ {t_{2\text{L}}} \right| $ (Fig.\ref{wave}(a1)) or $ \left|{t_{1\text{L}}}/ {t_{2\text{R}}} \right| < \left|{t_{1\text{R}}}/ {t_{2\text{L}}}\right| $(Fig.\ref{wave}(a3)). When $ \left|{t_{1\text{L}}} /{t_{2\text{R}}} \right| = \left|{t_{1\text{R}}} /{t_{2\text{L}}}\right|$(Fig.\ref{wave}(a2)), corresponding to the red imaginary  line in Fig.\ref{phase}(b), there is no skin effect and distribution of $|\psi_{a}\rangle$ and $|\psi_{b}\rangle$ is comparable.
In Fig.\ref{wave}(b), we show  $\chi$ versus $L$  both analytically and numerically with different values of $t_{1\text{L}}/t_{2\text{R}}$ by fixing the parameter $t_{1\text{R}}/t_{2\text{L}}=0.5$, which clearly indicates a transition from the right to left edge state.

\begin{figure}[htbp]
	\centering
	\includegraphics[width=0.8\columnwidth]{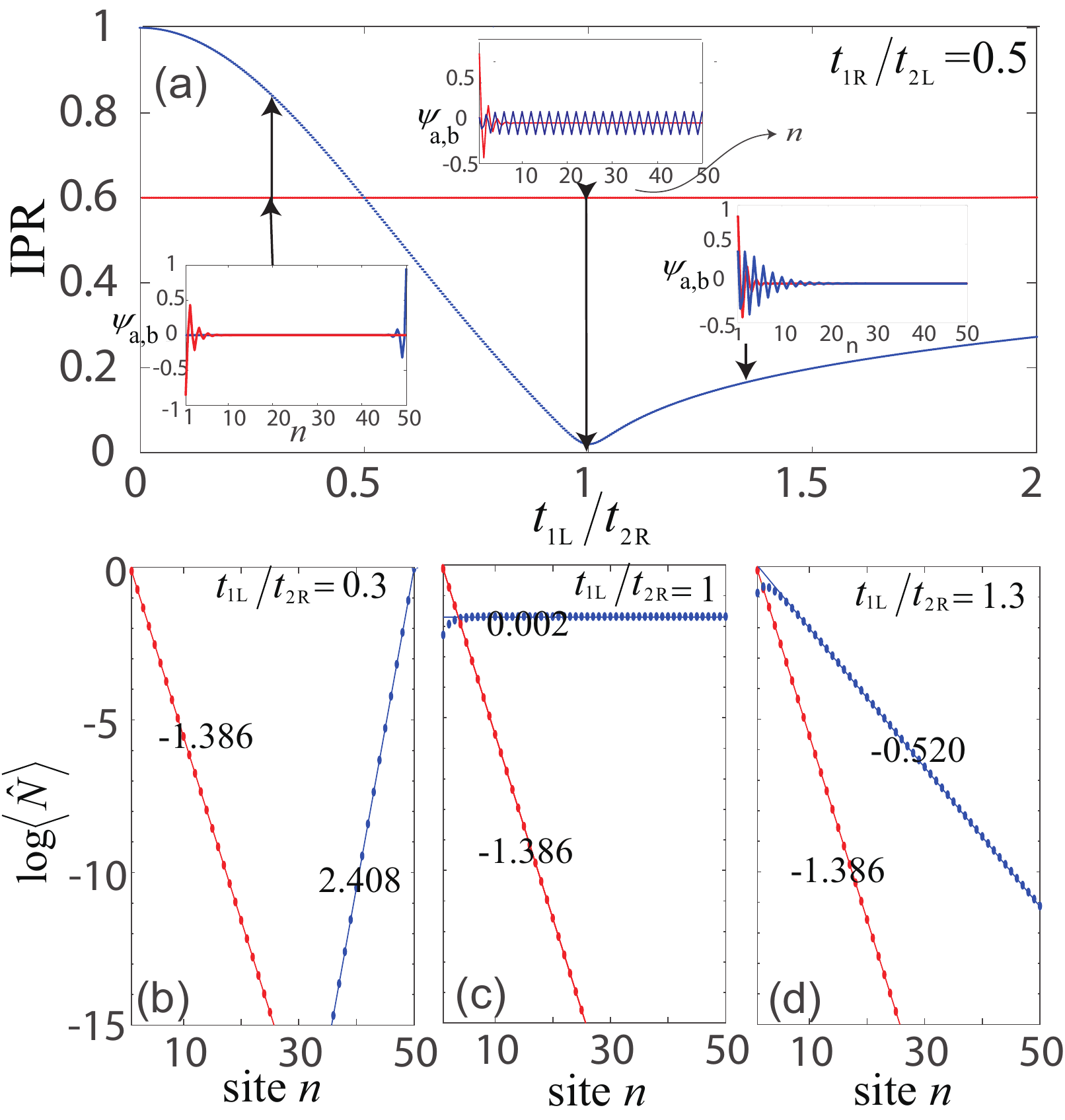}
\caption{(a) The IPR of two edge modes $|\psi_a\rangle$ and $|\psi_b\rangle$ with the red and blue curves corresponding to $|\psi_a\rangle$ and $|\psi_b\rangle$, respectively.
(b-d) The population distributions of edge modes with the red and blue dots corresponding to $|\psi_a\rangle$ and $|\psi_b\rangle$, respectively. The dots are the numerical results and the lines represent the theoretical fitting. The numbers marked in (b-d) represent the slop of numerical fitting. While the parameter $t_{1\text{R}}/t_{2\text{L}}=0.5$ is fixed, $t_{1\text{L}}/t_{2\text{R}}$ is marked in each figure. Here $t_{2\text{R}}=1$ and $t_{2\text{L}}=0.5$.}\label{IPR}
\end{figure}

Although $|\psi_{a,b}\rangle$ are always coupled together according to the finite size solutions, we can extract them from  $ |\psi_{1,2}\rangle $ via
$|\psi_{a}\rangle\propto |\psi_{1}\rangle+|\psi_{2}\rangle$ and $|\psi_{b}\rangle\propto |\psi_{1}\rangle-|\psi_{2}\rangle$.  To see how the modes of $|\psi_{a,b}\rangle$ changes with  $t_{1\text{L}}/t_{2\text{R}}$, we define the inverse participation ratios (IPR) for the modes (IPR$_{{a,b}}$) as
\begin{equation*}
  \text{IPR}_{i}=\sum_{n=1}^{L} \frac{|\langle n|\psi_{i}\rangle|^4}{|\langle \phi_{i}|\psi_{i}\rangle|^2},
\end{equation*}
where $i=a,b$ and $\langle n|=\langle n,A|+\langle n,B|$, and plot the IPR versus $t_{1\text{L}}/t_{2\text{R}}$ in Fig.\ref{IPR}(a) by fixing  $t_{1\text{R}}/t_{2\text{L}}=0.5$. While the value of IPR for an ideal localized state approaches $1$,  it approaches zero for an extended state.  When $t_{1\text{R}}/t_{2\text{L}}$ is fixed at $0.5$,  IPR of the mode $|\psi_{a}\rangle$ is not changed with $t_{1\text{L}}/t_{2\text{R}}$, with the corresponding wavefunction localized at the left edge. On the other hand, the IPR of the mode $|\psi_{b}\rangle$ displays a deep dive at $t_{1\text{L}}/t_{2\text{R}}=1$. As shown in Fig.\ref{IPR}(a), $|\psi_{b}\rangle$ undergoes a transition from the right edge state to left edge state, whereas the mode becomes an extended state at the transition point $t_{1\text{L}}/t_{2\text{R}}=1$. Such a transition can be also predicted by analyzing the analytical solutions.  In Fig.\ref{IPR}(b--d), we also display the population distribution $\hat {N}$, which is given by $\langle\hat {N}\rangle=\langle\psi_{i} |\hat {N}|\psi_{i}\rangle/\langle\psi_{i} |\psi_{i}\rangle$ where  $\hat {N}=|n,A\rangle\langle n,A|+|n,B\rangle\langle n,B|$, with $n=1,2,...,L$ and $i=a,b$.  For $|\psi_{a}\rangle$ and $|\psi_{b}\rangle$, the numerical results indicate that log$\langle\hat {N}\rangle/n$ are proportional to $2\text{log}|t_{1\text{R}}/t_{2\text{L}}|$ and $2\text{log}|t_{2\text{R}}/t_{1\text{L}}|$, respectively, as shown in Fig.\ref{IPR}(b-d), consistent with our analytical results.

\begin{figure}[htbp]
	\centering
	\includegraphics[width=0.8\columnwidth]{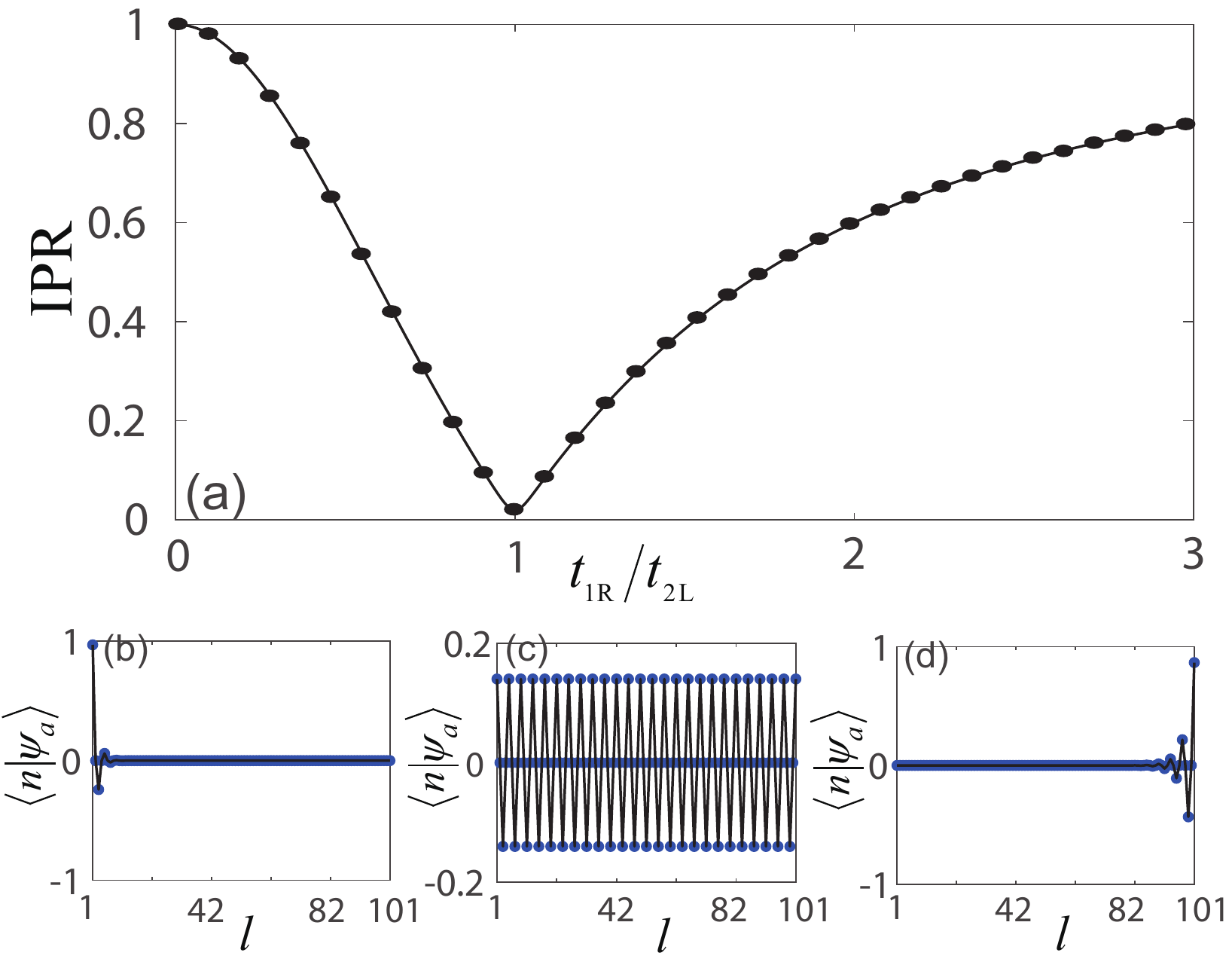}
\caption{(a)  The IPR of the edge state $|\psi_a\rangle$ for the open system terminated with $A$ site at both ends. %which is irrelevant to $t_{1\text{L}}/t_{2\text{R}}$
(b-d) represent the position distribution of $\langle n |\psi_a\rangle$. Parameters $(t_{1\text{R}},t_{2\text{L}})$  are (0.25,1) (b), (1,1) (c) and (2,1) (d).
The dots and lines represent the numerical and analytical results, respectively. }\label{y}
\end{figure}

According to the previous analysis, the symmetry $P$ ensures that the form of zero-mode states $|\psi_{1,2}\rangle$ in the finite-size system is always superposed by $|\psi_{a,b}\rangle$, and we can not solely observe the transition process of the mode $|\psi_{a,b}\rangle$. However, when the open system terminates with an $A~(B)$ site
at both ends, i.e. the total number of sites is odd and the symmetry $P$ is broken, we find that there is always a zero-mode state  $|\psi_{0}\rangle=|\psi_{a(b)}\rangle$ with energy $E=0$, whose wave function only distributes on the $A~(B)$ sublattice \cite{Kunst2019,Yuce2019}. We note that such a zero-mode state is inherently related to the breaking of the hidden P symmetry with no occurrence of finite size splitting.. According to Eq.(\ref{psia}), the zero-mode state  $|\psi_{a}\rangle$  distributes only on the sublattice $A$ and $\langle n|\psi_{a}\rangle$ is proportional to $(t_{1\text{R}}/t_{2\text{L}})^{n-1}$ , which suggests the distribution of  zero mode state would change from left to right edge when the parameter $t_{1\text{R}}/t_{2\text{L}}$ crosses over the the transition point $|t_{1\text{R}}/t_{2\text{L}}|=1$ from below. At the transition point, the zero mode wavefuntion would spread over all the lattice. This is verified by the numerical results as shown in Fig.\ref{y}, where the IPR of $|\psi_{a}\rangle$  takes a minimal value at $|t_{1\text{R}}/t_{2\text{L}}|=1$. Such an anomalous zero mode state at the transition point has also been numerically observed in Ref. \cite{Yuce2019}. Similarly, for the open system terminates with the $B$ site at both ends, we can observe a similar transition at $|t_{1\text{L}}/t_{2\text{R}}|=1$. An alternative study of the zero mode state under the OBC via similarity transformation can be found in Appendix D.

\section{Summary}
In summary, we studied a generalized non-reciprocal Su-Schrieffer-Heeger model and determined its phase diagram under both the periodical and open boundary conditions via the calculation of topological invariant and zero-mode edge states, respectively. The general geometrical meaning of topological invariants in momentum $k$ space is also discussed. We give two different interpretations in terms of winding number and linking properties of curves in the three-dimensional space, respectively. Taking account of  the normalization condition properly by using the biorthogonal eigenvectors,  we give analytically the condition for the occurrence of zero mode edge states under the OBC.
Then we construct explicitly the analytical expression of zero-mode edge states for the finite-size system by enforcing the wavefunctions to fulfill the hidden symmetry of the system. By using the analytical wavefunctions, we calculate the gap size of zero energy splitting and study the evolution of zero mode states. Our analytical results are found to agree very well with the numerical results via exact diagonalization of finite-size systems.

\begin{acknowledgments}
S. C. is supported by NSFC under Grants No. No.11974413 and the National Key Research and Development Program of China (2016YFA0300600 and 2016YFA0302104).
R. L. is supported by NSFC under Grants No. 11874234 and the National Key Research and Development Program of China (2018YFA0306504).
\end{acknowledgments}

{\bf  Author contribution statement}
\\
The project was supervised by S.C. H.J. carried out the calculations. H.J. and
S.C. wrote the manuscript. All authors provided critical feedback and helped shape the
research, analysis and manuscript.

\appendix
\appendix
\section{The topological invariance}\label{a}
In non-Hermitian system, the topological invariance related to the Berry phase can be defined as
\begin{equation*}
  \nu_{s,n}= \frac{1}{\pi} \int \text{d} {\bm{k}} \langle\phi_{n}|i\partial_{\bm{k}}|\psi _{n}\rangle,
\end{equation*}
where $|\psi(\phi)_n\rangle$ represents $n$-th right (left)-eigenvectors of $H(k)$ with $n$ being the band index.
The topological invariance $\nu_{s,1}$ can be represented as
\[
\nu_{s,1} = \frac{1}{\pi}\int \text{d} k A_{1,k},
\]
with
 \begin{eqnarray*}
            A_{1,k} &=& \langle\phi _{1}|i\partial_k|\psi _{1}\rangle, \\
             &=&\frac{1}{2}\left(
                             \begin{array}{cc}
                               {h_-}/{E_{1}} &1 \\
                             \end{array}
                           \right)i\partial_k\left(
                                               \begin{array}{c}
                                                {h_+}/{E_{1}} \\
                                                 1 \\
                                               \end{array}
                                             \right), \\
           &=&\frac{i}{4} \frac{h_-\partial_k h_+-h_+\partial_k h_- }{E^2_{1}},\\
           &=&\frac{i}{4} (\partial_k\text{In}h_+-\partial_k\text{In}h_-).
          \end{eqnarray*}
Since $h_{\pm}$ are generally complex, they can be written as $h_{\pm}=|h_{\pm}|e^{i\varphi_{\pm}}$, and thus the topological invariance $\nu_{s,1}$ is simplified to
\begin{eqnarray}
            \nu_{s,1} &=& \frac{1}{4\pi}\int \text{d} k (-\partial_k\varphi_++\partial_k\varphi_-),\\
                  &=&\frac{1}{2}(\nu_{-}-\nu_{+}).
          \end{eqnarray}
In the same way, the topological invariance $\nu_{s,2}$ is found to be equal to $\nu_{s,1}$, i.e., $(\nu_{s,2}=\nu_{s,1})$. For the winding number $\nu_E$ of eigenvalues $E_{1,2}$, it's easy to find  $\nu_E=(\nu_++\nu_-)/2$, i.e.,
\begin{eqnarray*}
  \nu_E &=&  \frac{1}{2\pi}\int \text{d} k \partial_k \text{Arg} (E_{1.2}) \\
   &=& \frac{1}{4\pi}\int \text{d} k \partial_k \text{Arg} (h_{+}\cdot h_-)\\
   &=& \frac{1}{4\pi}\int \text{d} k \partial_k (\varphi_{+}+\varphi_{-})  \\
   &=&  \frac{1}{2} (\nu_++\nu_-).
\end{eqnarray*}

It is clear both the topological invariance $\nu_{s,1(2)}$ and $\nu_E$ of the Hamiltonian with chiral symmetry can be split into  two parts $\nu_{+}$ and $\nu_{-}$, which denote the winding numbers of the trajectories of (Re$h_{\pm}(k)$, Im$h_{\pm}(k)$) encircling the
exceptional points $h_{+}=0$ and $h_{-}=0$, respectively. Similar to Ref.\cite{Lieu2018,Jiang2018}, the topological invariant $\nu_{\pm}$ can be written as
\begin{eqnarray*}
% \nonv_{\pm}umber to remove numbering (before each equation)
 \nu_{\pm}=\frac{1}{2}\sum_{i}(\mathrm{sgn}(\frac{\partial \mathrm{Im}(h_{\pm})}{\partial {k} }\mid_{{k}={K}_i})\cdot \mathrm{sgn}(\mathrm{Re}(h_{\pm})({K}_i)),
\end{eqnarray*}
with ${K}_i$ is the $i-$th solution of $\mathrm{Im}(h_{\pm})=0$. For the Hamiltonian described by Eq.(\ref{hk-SSH}), we fix parameters $t_{1(2)\text{R(L)}}$  to be real without loss of generality, and it is easy to get a simplified
form of $\nu_{\pm}$,
\begin{equation*}\label{nu}
  \begin{split}
  \nu_+&=\frac{1}{2}\{\mathrm{sgn}(t_{1\text{L}}- t_{2\text{R}})- \mathrm{sgn}(t_{1\text{L}}+t_{2\text{R}})\},\\
   \nu_-&=\frac{1}{2}\{\mathrm{sgn}(t_{1\text{R}}+ t_{2\text{L}})- \mathrm{sgn}(t_{1\text{R}}- t_{2\text{L}})\}.
  \end{split}
\end{equation*}

Since $h_{\pm}=h_{x}\pm ih_y$,
%($h_{x,y}=E_{i}\langle \sigma_{x,y} \rangle_i=E_{i}\langle \phi_i|\sigma_{x,y}|\psi_i \rangle$, $i=1,2$),
we can also represent $\varphi_{\pm}$ in terms of $h_x$ and $h_y$ as
\begin{eqnarray}
  \tan \varphi_{+} &=& - \frac{\text{Re}(h_y) - \text{Im}(h_x)}{\text{Re}(h_x)+\text{Im}(h_y)} ,\label{interwind1}\\
  \tan \varphi_{-}  &=& + \frac{\text{Re}(h_y)+\text{Im}(h_x)}{\text{Re}(h_x)-\text{Im}(h_y)} \label{interwind2},
\end{eqnarray}
similar to the definition in the previous reference \cite{Yin2018}. 
If we redefine $\varphi_{1}=\varphi_{-}$ and $\varphi_{2}=-\varphi_{+}$, we have
\begin{eqnarray*}
  \tan \varphi_{1}  &=&  \frac{\text{Re}(h_y)+\text{Im}(h_x)}{\text{Re}(h_x)-\text{Im}(h_y)}, \\
  \tan \varphi_{2} &=&  \frac{\text{Re}(h_y)-\text{Im}(h_x)}{\text{Re}(h_x)+\text{Im}(h_y)},
\end{eqnarray*}
which is identical to the definition in our previous work \cite{Yin2018}. Also, we have $\nu_{-}=\nu_{1}$ and $\nu_{+}=-\nu_2$, which lead to
\begin{eqnarray*}
 \nu_{s} &=&\frac{1}{2} (\nu_{1}+\nu_{2}), \\
 \nu_{E} &=& \frac{1}{2} (\nu_{1}-\nu_2) ,
\end{eqnarray*}
consistent with the previous work \cite{Yin2018}, where $\nu_{1,2}=\frac{1}{2\pi} \oint \partial_{k} \varphi_{1,2} \text{d} {k}$.\\

\begin{figure}[htbp]
	\centering
	\includegraphics[width=0.9\columnwidth]{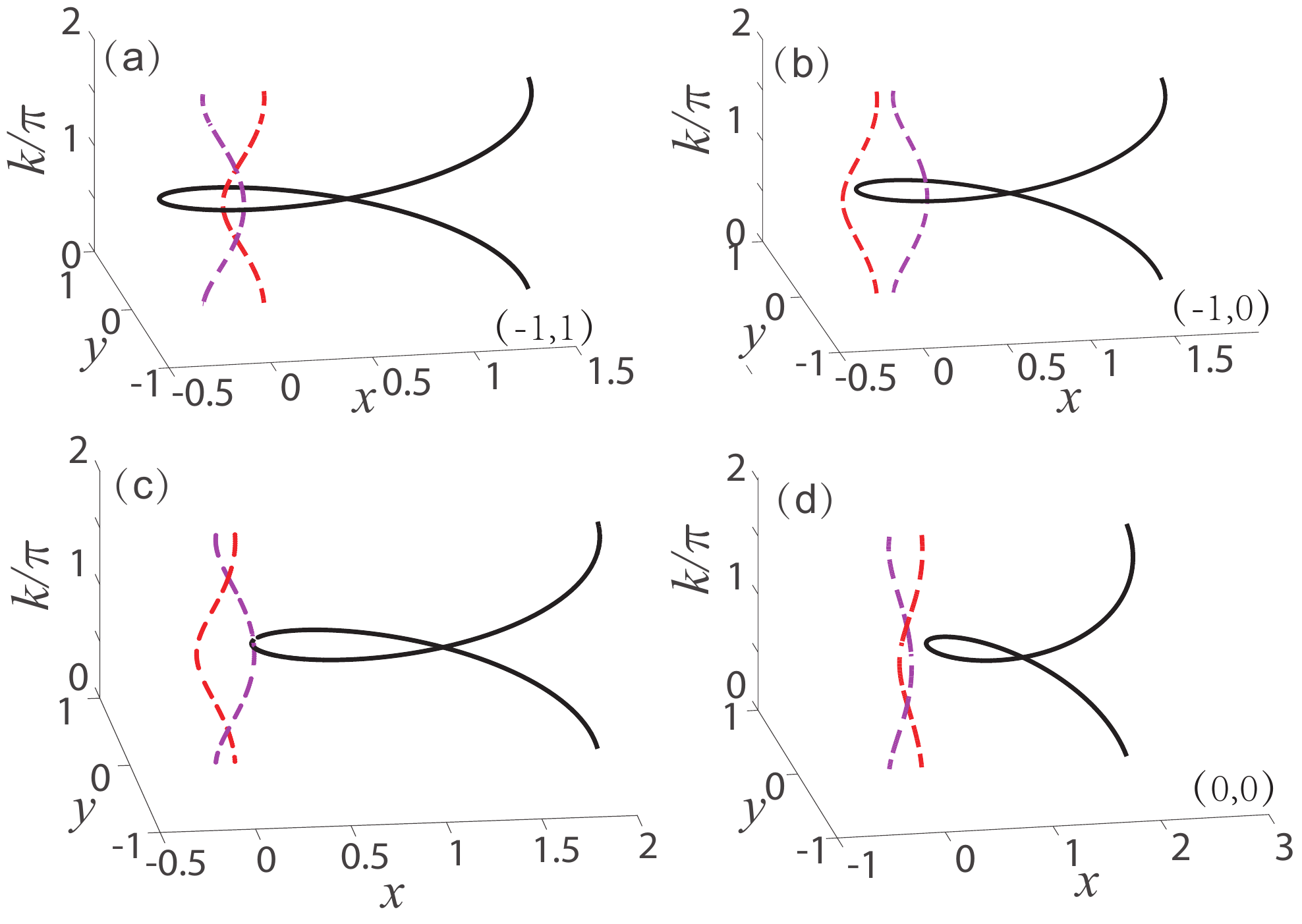}
\caption{Geometric configurations of `real' and `imaginary' curves demonstrate the linking properties of the curves can be described by the topological invariants $\nu_{\pm}$. While the black solid curve represents the trajectory of real part of Hamiltonian described by ($x$=Re$(h_{x}(k))$, $y$=Re$(h_{y}(k))$), the purple/red dashed curve denotes the trajectory of imaginary part described by ($x=\mp$Im$(h_y(k))$), $y=\pm$Im$(h_x(k))$). The corresponding winding number $(\nu_+,\nu_-)$ is marked on (a), (b) and (d), respectively. The parameters $(t_{1\text{L}},t_{1\text{R}},t_{2\text{L}},t_{2\text{R}})$ are (0.6,0.5,0.8,1) in (a), (0.6,0.9,0.8,1) in (b), (1,1.1,0.8,1) in (c) and
(1.2,1.1,0.8,1) in (d). The crossing point of  the `real' and `imaginary' curves in (c) is the exceptional points.}\label{yinl}
\end{figure}

\indent  When the momentum $k$ goes cross the Brillouin zone, the trajectory of the real part of Hamiltonian  forms a closed curve, which is described by
\begin{equation*}
\left\{
 \begin{array}{c}
               x=\text{Re}(h_{x}(k))  \\
               y=\text{Re}(h_{y}(k))
 \end{array}
\right. ,
\end{equation*}
as displayed in Fig.\ref{yinl} by the black curve. Similarly, we can plot the trajectories of the imaginary part of Hamiltonian described by
\begin{equation*}
\left\{
 \begin{array}{cc}
               x=-\text{Im}(h_{y}(k)) & \\
               y=+\text{Im}(h_{x}(k)) &
 \end{array}
 \right.   \text{and} ~~
 \left\{
\begin{array} {c}
               x=+\text{Im}(h_{y}(k)) \\
               y=-\text{Im}(h_{x}(k))
\end{array}
\right.   ,
\end{equation*}
%($x=\mp$Im$[h_{x}(k)]$, $y=\pm$Im$[h_{y}(k)]$),
which also form closed curves as shown in Fig.\ref{yinl} by the purple and red dashed curves, respectively.  It is shown that $\nu_{\pm}$ can describe the linking properties of two interwinding closed curves corresponding to trajectories of the real part and imaginary part of the Hamiltonian. Fig.\ref{yinl}(a), (b) and (d) correspond to the cases with the winding number $(\nu_+,\nu_-)=(-1,1)$, $(-1,0)$  and $(0,0)$, respectively. The phases with different winding numbers are topologically different and the interwinding curves can not transform continuously without crossing each other. As shown in Fig.\ref{yinl}(c), two closed curves cross at one of the exception point, at which $\text{Re}(h_x)=- \text{Im}(h_y)$ and $\text{Re}(h_y)=\text{Im}(h_x)$ or equivalently $h_{+}=0$. When $t_{2\text{L}} = t_{2\text{R}}$, the trajectories of the imaginary part of the Hamiltonian do not change with $k$, and it is convenient to project the trajectories into the two-dimensional space, consistent with the previous study.

%\indent According to the eigenvalue Eq.(\ref{eigenvalue}), the EPs satisfies $h_{+}=0$ or $h_{-}=0$. And EPs can be represented the origin of the coordinate (Re$(h_{\pm})$, Im$(h_{\pm})$)system, and the winding number $\nu_{\pm}$ is equal to the number of revolutions of (Re$(h_{\pm})$, Im$(h_{\pm}))$  around the origin, as shown like Fig.1(b). \\

\begin{figure}[htbp]
	\centering
	\includegraphics[width=0.9\columnwidth]{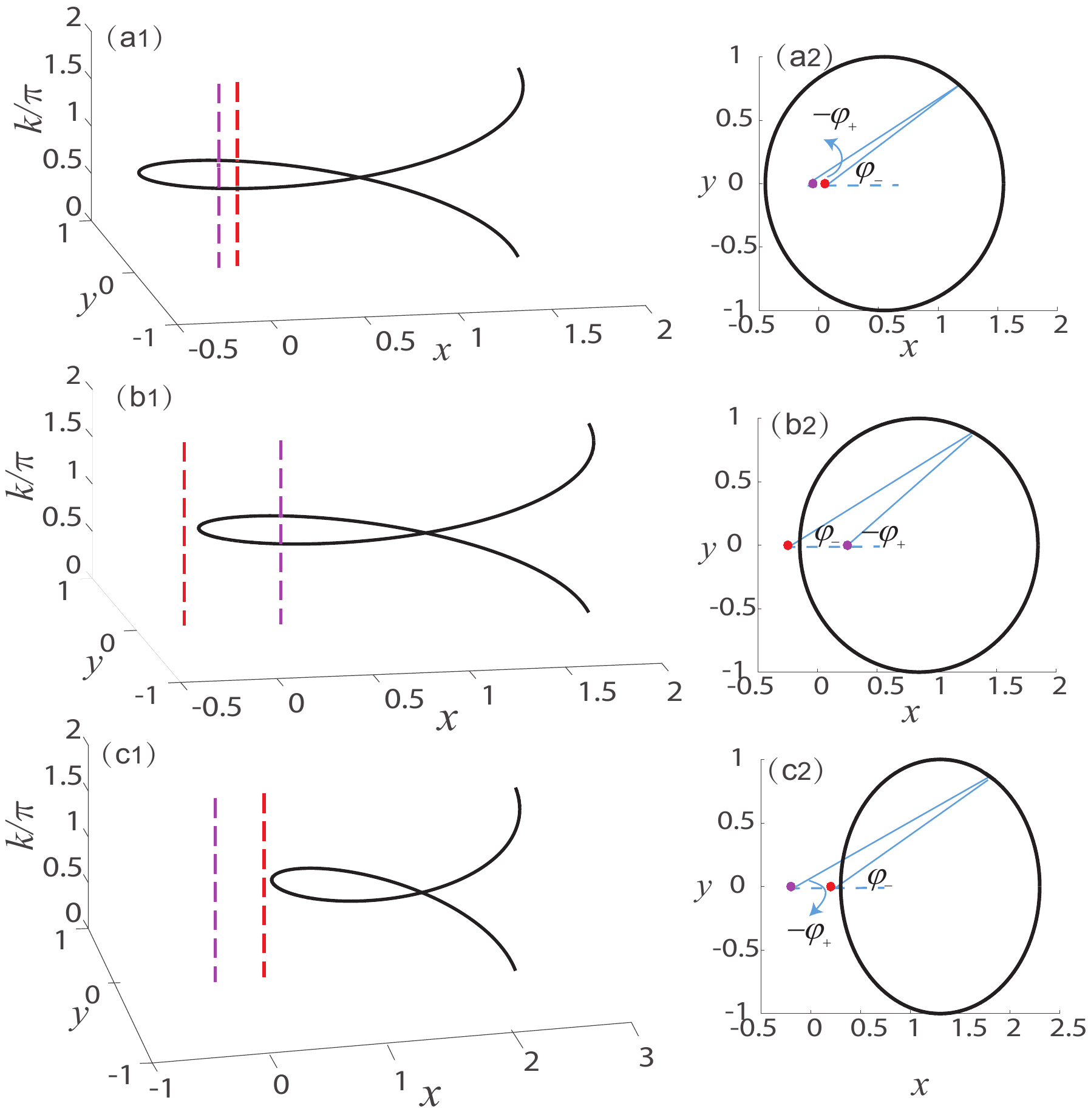}
\caption{ Geometric configurations of topological invariants $\nu_{\pm}$ for the system with parameters $t_{2\text{L}}=t_{2\text{R}}=1$.  While the black solid curve represents the trajectory of real part of Hamiltonian described by ($x$=Re$(h_{x}(k))$, $y$=Re$(h_{y}(k))$), the purple/red dashed curve denotes the trajectory of imaginary part described by ($x=\mp$Im$(h_y)$, $y=0$), which is independent of $k$. The  top views of (a1-c1) are shown in  (a2-c2). The other parameters $(t_{1\text{L}},t_{1\text{R}})$ are (0.6,0.5) in (a1,a2); (0.6,1.1) in (b1,b2); (1.5,1.1) in (c1,c2). Angles $(\varphi_+,\varphi_-)$ marked in (a2-c2) are  corresponding to the angles in Fig.1(b). }\label{yin2}
\end{figure}

As shown in Fig.\ref{yinl}, the linking of the  `real' curve  (Re$(h_{x}(k))$, Re$(h_{y}(k))$)with `imaginary' curves ($\mp\text{Im}(h_y(k))$,
    $\pm\text{Im}(h_x(k))$) gives a direct interpretation in terms of winding number $(\nu_+,\nu_-)$ or equivalently $(\nu_1,\nu_2)$.
    %To see this, the numbers ($n_1,n_2$) of entanglements of real curve and imaginary curves (purple, red) is in direct proportion to winding number $(\nu_+,\nu_-)$ with $\nu_+=-n_2,\nu_-=n_1$, as shown in  Fig.(\ref{yin2}).
When $t_{2\text{L}}=t_{2\text{R}}$, $ \text{Im}(h_{x,y})$ are independent of  the momentum $k$ and thus the `imaginary' curves ($\mp\text{Im}(h_y),\pm\text{Im}(h_x)$) become  two straight lines as shown in Fig.\ref{yin2} (a1)-(c1). The corresponding plane projections of the curves are shown in Fig.\ref{yin2} (a2)-(c2), demonstrating that the geometrical relationship about the `real' and `imaginary' curves can be well described by the winding number on the projected plane as in the previous work \cite{Yin2018}. However, for the general case with $t_{2\text{L}} \neq t_{2\text{R}}$,  since the imaginary parts of Hamiltonian $ \text{Im}(h_{x,y})$  are functions of momentum $k$,  the plane projections of `imaginary' curves form also closed curve, and  there may exit `fake' intersection between the projected `real' and `imaginary' curves although they do not interwind  in the three-dimensional space. So it is more natural to see the linking properties of the `real' and  `imaginary' curves in the  three-dimensional parameter space to describe the topological properties of the general non-reciprocal two-band systems with chiral symmetry.

\section{zero mode edge state in the semi-infinite system}\label{semi}
 Consider the non-Hermitian lattice model
 \begin{equation}
    \tilde{\mathcal{H}}=\text{I}\otimes\left(
                             \begin{array}{cc}
                               0 &t_{1\text{L}} \\
                               t_{1\text{R}} & 0 \\
                             \end{array}
                           \right)
                           +\hat{S}\otimes\left(
                             \begin{array}{cc}
                               0 & t_{2\text{R}}\\
                               0 & 0 \\
                             \end{array}
                           \right)
                           +\hat{S}^{\dagger}\otimes\left(
                             \begin{array}{cc}
                               0 & 0 \\
                              t_{2\text{L}} & 0 \\
                             \end{array}
                           \right),
  \end{equation}
  with  unit operator I, backward and forward translation
operators $\hat{S}|i\rangle=|i+1\rangle,\hat{S}^{\dagger}|i\rangle=|i-1\rangle$. As $S^{\dagger}_{i,j}=\langle i|\hat{S}|j\rangle=\delta_{i,j+1}$ and $S^{\dagger}_{i,j}=\langle i|\hat{S}^{\dagger}|j\rangle=\delta_{i,j-1}$. The corresponding matrices are,
  \begin{equation*}
\centering {\begin{matrix}
S=\begin{pmatrix}
   0&0&&&&\\
    1&0&0&&&\\
    &1&0&0&&\\
    &&1&0&\ddots&\\
    &&&\ddots&\ddots&\\
  \end{pmatrix};\quad   S^{\dagger}=\begin{pmatrix}
   0&1&&&&\\
    0&0&1&&&\\
    &0&0&1&&\\
    &&0&0&\ddots&\\
    &&&\ddots&\ddots&\\
  \end{pmatrix}
\end{matrix}}.
\end{equation*}
with dimension $L$. The system is divided into two subsystems: position $n=(1,2,..L)$ and sublattice $A,B$. \\
\indent Using the ansatz  $|\tilde{\psi}\rangle=1/\mathcal{N}\sum_{n=1}^{L}\beta^{n-1}|n\rangle\otimes|\xi\rangle$,$|\tilde{\phi}\rangle=1/\mathcal{N}\sum_{n=1}^{L}\beta'^{n-1}|n\rangle\otimes|\xi\rangle$  with normalization constant $\mathcal{N}=\sqrt{(1-(\beta'\beta)^L)/(1-\beta'\beta)}$ and the information of sublattice A/B, $|\xi\rangle$ £¬ we have  $ \tilde{\mathcal{H}}|\tilde{\psi}\rangle=E|\tilde{\psi}\rangle$ , $ \tilde{\mathcal{H}}^{\dagger}|\tilde{\phi}\rangle=E^*|\tilde{\phi}\rangle$ and $\langle\tilde{\phi}|\tilde{\psi}\rangle=1$, the Schr\"{o}dinger equation for the real space system leads the relations for the bulk $(1<n<L)$:
\begin{equation}\label{bulk}
  \left(
                             \begin{array}{cc}
                               0 &t_{1\text{L}}+t_{2\text{R}}\beta^{-1} \\
                               t_{1\text{R}}+t_{2\text{L}}\beta & 0 \\
                             \end{array}
                           \right)|\xi\rangle=E|\xi\rangle,
\end{equation}
where $E^2={(t_{1\text{L}}+t_{2\text{R}}\beta^{-1})( t_{1\text{R}}+t_{2\text{L}}\beta )}$. And for the boundary at $ n = 1$  ,we have
\begin{equation}\label{n=1}
  \left(
                             \begin{array}{cc}
                               0 &t_{1\text{L}} \\
                                t_{1\text{R}}+t_{2\text{L}}\beta & 0 \\
                             \end{array}
                           \right)|\xi\rangle=E|\xi\rangle,
\end{equation}
where $E^2={t_{1\text{L}}(  t_{1\text{R}}+t_{2\text{L}}\beta)}$. And the energy of Eq.(\ref{bulk},\ref{n=1})  need to be consistent with each other. Comparing the difference between these two relations $n$ ($1<n<L$) Eq.(\ref{bulk}) and $n=1$ Eq.(\ref{n=1}), we can get the zero mode state
\begin{equation*}
  |{\psi}_a\rangle=1/\mathcal{N}_a\sum_{n=1}^{N_a}\beta_a^{n-1}|n\rangle\otimes|\xi_a\rangle,
\end{equation*}
with $|\xi_a\rangle=(1,0)^\text{T}$, $\beta_a=-t_{1\text{R}}/t_{2\text{L}}$, $\beta'_a=-t_{1\text{L}}/t_{2\text{R}}$ and $\mathcal{N}_a=\sqrt{(1-(\beta'_a\beta_a)^L)/(1-\beta'_a\beta_a)}$  under the condition of  $|t_{1\text{L}}t_{1\text{R}}|<|t_{2\text{L}}t_{2\text{R}}|$ and the length $L\rightarrow\infty$.\\

In the same way,  considering the semi-infinite limit from the right boundary, we have the right zero-mode state of the following form:
\begin{equation}
  |{\psi}_b\rangle=1/\mathcal{N}_b\sum_{n=0}^{L-1}\beta_b^{n}|L-n\rangle\otimes|\xi_b\rangle ,    \label{psi-b}
\end{equation}
with $|\xi_b\rangle=(0,1)^\text{T}$ and $\beta_b=- t_{1\text{L}}/t_{2\text{R}}=\beta'_a$
and $\mathcal{N}_b=\mathcal{N}_a$ under the condition of  $|t_{1\text{L}}t_{1\text{R}}|<|t_{2\text{L}}t_{2\text{R}}|$ and the length $L\rightarrow\infty$.
When the semi-infinite boundary condition is considered, we get two zero-mode states distributing only on the A (or B) sublattice, according to $\beta_a=-t_{1\text{R}}/t_{2\text{L}}$ and $\beta_b=-t_{1\text{L}}/t_{2\text{R}}$, respectively.
When $|\beta_a|=|\beta_b|^{-1}$ ($|t_{1\text{L}}t_{1\text{R}}|=|t_{2\text{L}}t_{2\text{R}}|$), the system has the transition point. Considering the constraints, only when  $|t_{1\text{L}}t_{1\text{R}}|<|t_{2\text{L}}t_{2\text{R}}|$ and the length of the system $L\rightarrow\infty$, the system has  zero mode states $|\psi_{a,b}\rangle$. And for independent zero mode states $|\psi_{a,b}\rangle$, any linear combination of $|\psi_{a,b}\rangle$  is still the zero mode solution of this Hamiltonian (Eq.(\ref{OBC1})).

\section{A hidden symmetry of the system under the OBC}\label{C11}
The Hamiltonian under the OBC can be rewritten as
\begin{equation*}\label{OBC}
\begin{split}
   & \tilde{\mathcal{H}}=\sum_{n=1}|n\rangle\langle n|\otimes\left(
                             \begin{array}{cc}
                               0 &t_{1\text{L}} \\
                               t_{1\text{R}} & 0 \\
                             \end{array}
                           \right)+
                            |n\rangle\langle n-1|\otimes\left(
                             \begin{array}{cc}
                               0 & t_{2\text{R}}\\
                               0 & 0 \\
                             \end{array}
                           \right)\\
                           &\qquad\qquad\qquad\qquad+
                           |n\rangle\langle n+1|\otimes\left(
                             \begin{array}{cc}
                               0 & 0 \\
                              t_{2\text{L}} & 0 \\
                             \end{array}
                           \right),
    \end{split}
  \end{equation*}
identical to Eq.(\ref{OBC1}). It is noticed that there exists a hidden symmetry for the non-Hermitian Hamiltonian. Given the operator $P$, defined by
 \begin{equation*}
\begin{split}
P&=\sum_{n=1}^{L} r^{L-2n+1}|L-n+1\rangle\langle n|\otimes \left(
                                                               \begin{array}{cc}
                                                                 0 & \alpha^{-1} \\
                                                                  \alpha & 0 \\
                                                               \end{array}
                                                             \right).\\
                                                             &
                                                             \end{split}
\end{equation*}\\
with $r=\sqrt{t_{1\text{R}}t_{2\text{R}}/t_{1\text{L}}t_{2\text{L}}}$ and $\alpha=\sqrt{t_{1\text{R}}/t_{1\text{L}}}$,  we can see that the operator $P$ satisfies
  \begin{eqnarray*}
   \tiny  P^2 &=& \sum_{n,m} r^{L-2n+1}|L-n+1\rangle\langle n|\otimes \left(
                                                               \begin{array}{cc}
                                                                 0 & \alpha^{-1} \\
                                                                  \alpha & 0 \\
                                                               \end{array}
                                                             \right)\\
          &&    \quad\qquad\quad         \times  r^{L-2m+1}|L-m+1\rangle\langle m|\otimes \left(
                                                               \begin{array}{cc}
                                                                 0 & \alpha^{-1} \\
                                                                  \alpha & 0 \\
                                                               \end{array}
                                                             \right) \\
     &=& \sum_{n,m}  r^{2L-2n-2m+2} |L-n+1\rangle\langle n|L-m+1\rangle\langle m|\otimes   \left(
                                                               \begin{array}{cc}
                                                                 1 & 0 \\
                                                                  0 & 1 \\
                                                               \end{array}
                                                             \right) \\
     &=&  \sum_{n}   |L-n+1\rangle\langle L-n+1 | \otimes   \left(
                                                               \begin{array}{cc}
                                                                 1 & 0 \\
                                                                  0 & 1 \\
                                                               \end{array}
                                                             \right) \\
     &=&\text{ unit matrix}.
  \end{eqnarray*}
  Also, we have $ P^{\dagger} \neq  P$. Next we prove that the Hamiltonian satisfies $P \tilde{\mathcal{H}} P=\tilde{\mathcal{H}}$ as follows: \\
 \begin{small}
 \begin{widetext}
  \begin{eqnarray*}
   P \tilde{\mathcal{H}} P&=&  \sum_{n=1}^{L} r^{L-2n+1}|L-n+1\rangle\langle n|\otimes \left(
                                                               \begin{array}{cc}
                                                                 0 & \alpha^{-1} \\
                                                                  \alpha & 0 \\
                                                               \end{array}
                                                             \right) \\
   && \qquad\qquad \times\left[\sum_{n'=1}^{L}|n'\rangle\langle n'|\otimes\left(
                             \begin{array}{cc}
                               0 &t_{1\text{L}} \\
                               t_{1\text{R}} & 0 \\
                             \end{array}
                           \right)+
                           \sum_{n'=2}^{L}|n'\rangle\langle n'-1|\otimes\left(
                             \begin{array}{cc}
                               0 & t_{2\text{R}}\\
                               0 & 0 \\
                             \end{array}
                           \right)
                           +
                            \sum_{n'=1}^{L-1}|n'\rangle\langle n'+1|\otimes\left(
                             \begin{array}{cc}
                               0 & 0 \\
                              t_{2\text{L}} & 0 \\
                             \end{array}
                           \right)\right]\\
   &&  \qquad\qquad\qquad\qquad\qquad\qquad\qquad\qquad\qquad\qquad\times\sum_{m=1}^{L} r^{L-2m+1}|L-m+1\rangle\langle m|\otimes \left(
                                                               \begin{array}{cc}
                                                                 0 & \alpha^{-1} \\
                                                                  \alpha & 0 \\
                                                               \end{array}
                                                             \right)\\
      &=&         \sum_{n=1}^{L}|n\rangle\langle n|\otimes\left(
                             \begin{array}{cc}
                               0 &t_{1\text{L}} \\
                               t_{1\text{R}} & 0 \\
                             \end{array}
                           \right)+
                            \sum_{n=2}^{L}  |L-n+1\rangle\langle L-n+2 |\otimes\left(
                             \begin{array}{cc}
                               0 & 0 \\
                              t_{2\text{L}} & 0 \\
                             \end{array}
                           \right) +
                           \sum_{n=0}^{L-1}
                           |L-n+1\rangle\langle L-n|\otimes\left(
                             \begin{array}{cc}
                               0 & t_{2\text{R}}\\
                               0 & 0 \\
                             \end{array}
                           \right)          \\
      &=&         \sum_{m=1}^{L}|m\rangle\langle m|\otimes\left(
                             \begin{array}{cc}
                               0 &t_{1\text{L}} \\
                               t_{1\text{R}} & 0 \\
                             \end{array}
                           \right)+
                            \sum_{m=1}^{L-1} |m\rangle\langle m+1 |\otimes\left(
                             \begin{array}{cc}
                               0 & 0 \\
                              t_{2\text{L}} & 0 \\
                             \end{array}
                           \right) +
                           \sum_{m=2}^{L}
                           |m\rangle\langle m-1|\otimes\left(
                             \begin{array}{cc}
                               0 & t_{2\text{R}}\\
                               0 & 0 \\
                             \end{array}
                           \right)\\
      &=& \tilde{\mathcal{H}}.
\end{eqnarray*}
\end{widetext}
 \end{small}
The zero mode states of semi-infinite Hamiltonian are
\begin{equation*}
  |{\psi}_a\rangle=1/\mathcal{N}_a\sum_{n=1}^{L-1}\beta_a^{n-1}|n\rangle\otimes|\xi_a\rangle,
\end{equation*}
with $|\xi_a\rangle=(1,0)^\text{T}$ and $\beta_a=-t_{1\text{R}}/t_{2\text{L}}$,
and
\begin{equation*}
  |{\psi}_b\rangle=1/\mathcal{N}_a\sum_{n=0}^{L-1}\beta_b^{n}|L-n\rangle\otimes|\xi_b\rangle,
\end{equation*}
with $|\xi_b\rangle=(0,1)^\text{T}$ and $\beta_b=- t_{1\text{L}}/t_{2\text{R}}=\beta'_a$.
It is straightforward to get
\begin{widetext}
\begin{eqnarray*}
   P |{\psi}_a\rangle &=& \sum_{n=1}^{L} r^{L-2n+1}|L-n+1\rangle\langle n|\otimes \left(
                                                               \begin{array}{cc}
                                                                 0 & \alpha^{-1} \\
                                                                  \alpha & 0 \\
                                                               \end{array}
                                                             \right)\cdot \left( 1/\mathcal{N}_a\sum_{n'=1}^{L-1}\beta_a^{n'-1}|n'\rangle\otimes|\xi_a\rangle\right)\\
                  &=&\alpha/\mathcal{N}_a \sum_{n,n'}r^{L-2n+1}\beta_a^{n'-1}    |L-n+1\rangle \delta_{n,n'}    \otimes        |\xi_b\rangle\\
                  &=& \alpha /\mathcal{N}_a  \sum_{n}    r^{L-2n+1}\beta_a^{n-1}    |L-n+1\rangle \otimes|\xi_b\rangle\\
                  &=& \alpha/ \mathcal{N}_a  \sum_{n}    {\sqrt{\beta_a/\beta_b}}^{L-2n+1}\beta_a^{n-1}    |L-n+1\rangle \otimes|\xi_b\rangle\\
                   &=& \alpha/ \mathcal{N}_a  \sum_{n}   {r}^{L-1}  \beta_b^{n-1}    |L-n+1\rangle \otimes|\xi_b\rangle\\
                   &=& {r}^{L-1}  \cdot \alpha/ \mathcal{N}_a  \sum_{m=0}^{L-1}  \beta_b^{m}    |L-m\rangle \otimes|\xi_b\rangle\\
                 &=& \alpha \cdot {r}^{L-1}|{\psi}_b\rangle.
  \end{eqnarray*}
  \end{widetext}
In the same way, we have:
  \begin{eqnarray*}
    P |{\psi}_b\rangle &=& \frac{\alpha^{-1}}{{r}^{L-1}}\cdot|{\psi}_a\rangle.
  \end{eqnarray*}
As a result, the operator $P$ satisfies  $P |{\psi}_a\rangle=\alpha \cdot {r}^{L-1}|{\psi}_b\rangle$ and  $P (\alpha \cdot {r}^{L-1}|{\psi}_b\rangle)= |{\psi}_a\rangle$ . Given
\begin{eqnarray}
|{\psi}_1\rangle&=&  |{\psi}_a\rangle+\alpha \cdot {r}^{L-1}|{\psi}_b\rangle,\\
 |{\psi}_2\rangle&=&  |{\psi}_a\rangle-\alpha \cdot {r}^{L-1}|{\psi}_b\rangle,
\end{eqnarray}
with $\alpha=\sqrt{t_{1\text{R}}/t_{1\text{L}}}$ and $r=\sqrt{t_{1\text{R}}t_{2\text{L}}/t_{1\text{L}}t_{2\text{R}}}$, we can obtain
\begin{eqnarray}
 P |{\psi}_1\rangle &=& |{\psi}_1\rangle,\\
  P |{\psi}_2\rangle&=& -|{\psi}_2\rangle.
\end{eqnarray}

By using Eq.(\ref{phi}), it is straightforward to calculate\begin{eqnarray*}
                                                              E_{1(2)}&=& \langle\phi_{1(2)}|\tilde{\mathcal{H}}|\psi_{1(2)}\rangle, \\
                                                               &=& \frac{1}{{\langle\phi_{1(2)}|\psi_{1(2)}\rangle}}(r^{{L-1}}\cdot\alpha\langle\phi_{a}|+(-)\langle\phi_{b})|\\                                                     &&\quad\qquad\quad\qquad\quad\cdot \tilde{\mathcal{H}}( |\psi_{a}\rangle+(-) r^{{L-1}}\cdot\alpha|\psi_{b}\rangle)
                                                            \end{eqnarray*}
where $\langle{\phi}_a| \tilde{\mathcal{H}} |{\psi}_a\rangle = \langle{\phi}_b| \tilde{\mathcal{H}} |{\psi}_b\rangle=0$, ${\langle\phi_{1(2)}|\psi_{1(2)}\rangle}=2\alpha r^{{L-1}}$ and
 \begin{eqnarray*}
   \langle{\phi}_b| \tilde{\mathcal{H}} |{\psi}_a\rangle&=& t_{1\text{R}}\beta_a^{L-1}/\mathcal{N}_a^2, \\
   \langle{\phi}_a| \tilde{\mathcal{H}}|{\psi}_b\rangle&=& t_{1\text{L}}\beta_b^{L-1}/\mathcal{N}_a^2.
 \end{eqnarray*}
The specific calculation process is as follows:
  \begin{widetext}
  \begin{eqnarray*}
    \langle{\phi}_b| \tilde{\mathcal{H}} |{\psi}_a\rangle&=&  \langle{\phi}_b|\left[\sum_{n'=1}^{L}|n'\rangle\langle n'|\otimes\left(
                             \begin{array}{cc}
                               0 &t_{1\text{L}} \\
                               t_{1\text{R}} & 0 \\
                             \end{array}
                           \right)+
                           \sum_{n'=2}^{L}|n'\rangle\langle n'-1|\otimes\left(
                             \begin{array}{cc}
                               0 & t_{2\text{R}}\\
                               0 & 0 \\
                             \end{array}
                           \right)
                           +
                            \sum_{n'=1}^{L-1}|n'\rangle\langle n'+1|\otimes\left(
                             \begin{array}{cc}
                               0 & 0 \\
                              t_{2\text{L}} & 0 \\
                             \end{array}
                           \right)\right]\\
   &&  \qquad\qquad\qquad\qquad\qquad\qquad\qquad\qquad\qquad\qquad\cdot \left( 1/\mathcal{N}_a\sum_{n=1}^{L-1}\beta_a^{n-1}|n\rangle\otimes|\xi_a\rangle\right)\\
      &=&       \langle{\phi}_b| 1/\mathcal{N}_a   \sum_{n}t_{1\text{R}}\beta_a^{n-1}|n\rangle\otimes|\xi_b\rangle +t_{2\text{L}} \beta_a^{n-1}|n-1\rangle\otimes|\xi_b\rangle\\
      &=&        (1/\mathcal{N}_a\sum_{m}{\beta^{m}_a}\langle L-m|\otimes\langle\xi_b|)\cdot(1/\mathcal{N}_a   \sum_{n}t_{1\text{R}}\beta_a^{n-1}|n\rangle\otimes|\xi_b\rangle +t_{2\text{L}} \beta_a^{n-1}|n-1\rangle\otimes|\xi_b\rangle\\  \\
      &=&\beta_a^{L-1}/\mathcal{N}_a^2( t_{1\text{R}}L-t_{1\text{R}}(L-1))\\
      &=&t_{1\text{R}}\beta_a^{L-1}/\mathcal{N}_a^2,
\end{eqnarray*}
\end{widetext}

  \begin{widetext}
  \begin{eqnarray*}
    \langle{\phi}_a| \tilde{\mathcal{H}} |{\psi}_b\rangle&=&  \langle{\phi}_a|\left[\sum_{n'=1}^{L}|n'\rangle\langle n'|\otimes\left(
                             \begin{array}{cc}
                               0 &t_{1\text{L}} \\
                               t_{1\text{R}} & 0 \\
                             \end{array}
                           \right)+
                           \sum_{n'=2}^{L}|n'\rangle\langle n'-1|\otimes\left(
                             \begin{array}{cc}
                               0 & t_{2\text{R}}\\
                               0 & 0 \\
                             \end{array}
                           \right)
                           +
                            \sum_{n'=1}^{L-1}|n'\rangle\langle n'+1|\otimes\left(
                             \begin{array}{cc}
                               0 & 0 \\
                              t_{2\text{L}} & 0 \\
                             \end{array}
                           \right)\right]\\
   &&  \qquad\qquad\qquad\qquad\qquad\qquad\qquad\qquad\qquad\qquad\cdot \left( 1/\mathcal{N}_a\sum_{n=1}^{L-1}\beta_b^{n}|L-n\rangle\otimes|\xi_b\rangle\right)\\
      &=&       \langle{\phi}_b| 1/\mathcal{N}_a   \sum_{n}t_{1\text{L}}\beta_b^{n}|L-n\rangle\otimes|\xi_a\rangle +t_{2\text{R}} \beta_b^{n}|L-n+1\rangle\otimes|\xi_a\rangle\\
      &=&        (1/\mathcal{N}_a\sum_{m}{\beta^{m-1}_b}\langle m|\otimes\langle\xi_b|)\cdot( 1/\mathcal{N}_a   \sum_{n}t_{1\text{L}}\beta_b^{n}|L-n\rangle\otimes|\xi_a\rangle +t_{2\text{R}} \beta_b^{n}|L-n+1\rangle\otimes|\xi_a\rangle )  \\
      &=& t_{1\text{L}}\beta_b^{L-1}/\mathcal{N}^2_a(L-(L-1))\\
      &=& t_{1\text{L}}\beta_b^{L-1}/\mathcal{N}^2_a.
\end{eqnarray*}
\end{widetext}
With simplification, we get
\begin{eqnarray*}
  E_1= &=& \frac{1}{2r^{{L-1}}\cdot\alpha}(r^{2L-2}\cdot\alpha^2 \langle{\phi}_a| \tilde{\mathcal{H}} |{\psi}_b\rangle+ \langle{\phi}_b| \tilde{\mathcal{H}} |{\psi}_a\rangle   ) \\
   &=&   \frac{1}{r^{{L-1}}\cdot\alpha}(t_{1\text{R}}\beta_a^{L-1}/\mathcal{N}^2_a  )\\
    &=&   \sqrt{t_{1\text{L}}t_{1\text{R}}}\sqrt{\beta_a\beta_b}^{L-1}/\mathcal{N}^2_a \\
    &=&\frac{\sqrt{t_{1\text{L}}t_{1\text{R}}}}{\mathcal{N}^2_a }\sqrt{\frac{t_{1\text{L}}t_{1\text{R}}}{t_{2\text{L}}t_{2\text{R}}}}^{L-1}.
\end{eqnarray*}
In the same way, we can obtain the $  E_2= -E_1 $. So the energy splitting at finite size $L$ is given by $\Delta E =E_1-E_2$, which reads
  \begin{equation}\label{deviation}
\begin{split}
  \Delta E&=\frac{2\sqrt{t_{1\text{L}}t_{1\text{R}}}}{\mathcal{N}^2_a }\sqrt{\frac{t_{1\text{L}}t_{1\text{R}}}{t_{2\text{L}}t_{2\text{R}}}}^{L-1}.
  \end{split}
\end{equation}

\section{Solution of non-Hermitian model under the OBC via similarity transformation}
 To understand the bulk states  in the non-Hermitian SSH model,
 the non-Hermitian matrix $\tilde{\mathcal{H}}$ can be transformed to a Hermitian one via a similarity transformation $V=\rho\otimes\rho_s$ ,
 \begin{eqnarray*}
    H'  &=& V^{-1} \tilde{\mathcal{H}} V, \\
   &=&  \text{I}\otimes \sqrt{ t_{1\text{L}}t_{1\text{R}}} \sigma_x +\frac{1}{2}(S+S^{\dagger})\otimes \sqrt{ t_{2\text{L}}t_{2\text{R}}} \sigma_x\\
   && \quad\qquad \qquad +\frac{1}{2i}(S-S^{\dagger})\otimes \sqrt{ t_{2\text{L}}t_{2\text{R}}} \sigma_y,
 \end{eqnarray*}
   where
             \begin{equation*}
             \begin{split}
               V&=\mathrm{diag}(1,r,r^2,....)\otimes\left(
                                                      \begin{array}{cc}
                                                        1 & 0 \\
                                                        0 & \sqrt{{t_{1\text{R}}}/{t_{1\text{L}}}} \\
                                                      \end{array}
                                                    \right),\\
                \end{split}
             \end{equation*}
with $r=\sqrt{t_{1\text{R}}t_{2\text{R}}/t_{1\text{L}}t_{2\text{L}}}$.

   \indent For the Hermitian Hamiltonian $H'$, the eigenvectors $|\psi'_m\rangle$ fulfill $H'|\psi'_m\rangle =E_m\psi'_m$ ($m=1,2,..2L$). They are related to the non-Hermitian eigenvectors via $|\psi_m\rangle=V|\psi'_m\rangle$,  where $\tilde{\mathcal{H}}|\psi_m\rangle=E_m |\psi_m\rangle$.  While the bulk state of the Hermitian case has a small value of IPR, the skin phase should have a larger value of IPR due to its boundary-localization nature (similarity transformation $V$).
%The distribution of quantity $\langle\hat{N}\rangle_m$ can describe the distribution of states which is exponentially decaying from one boundary, and the log$\langle\hat {N}\rangle$ is in direct proportion to $2\text{log}  |r|\cdot n $ for bulk states of non-Hermitian $\tilde{\mathcal{H}}$ .\\

Alternatively, we can also understand the fate of zero-mode edge states in the scheme of the similarity transformation.
 For a Hermitian SSH model, it is known that there exist zero-energy edge states only when $|t_{1\text{L}}t_{1\text{R}}|<|t_{2\text{L}}t_{2\text{R}}|$, the zero mode edge state $|\psi'_{a,b}\rangle$ with $\beta''_a=-\sqrt{ t_{1\text{L}} t_{1\text{R}}/t_{2\text{L}} t_{2\text{R}}}=\beta''_b$, whereupon the  $|\psi_{a,b}\rangle=V|\psi'_{a,b}\rangle$ is exponentially decaying from one boundary with ratio $\beta_a=-t_{1\text{R}}/t_{2\text{L}}$, $\beta_b=-t_{1\text{L}}/t_{2\text{R}}$. And the `zero' mode states could be written as $|\psi'_{1,2}\rangle\propto|\psi'_{a}\rangle\pm|\psi'_{b}\rangle$, the ratio $\chi'=1$, then it's easy to obtain $|\psi_{1,2}\rangle=V|\psi'_{1,2}\rangle$ and  ratio $\chi=\sqrt{{t_{1\text{L}}}/{t_{1\text{R}}}}\cdot (1/r)^{L-1}$. Comparing with the analysis in the main text, the different methods give the same results. \\
   \begin{figure}[htbp]
	\centering
	\includegraphics[width=0.9\columnwidth]{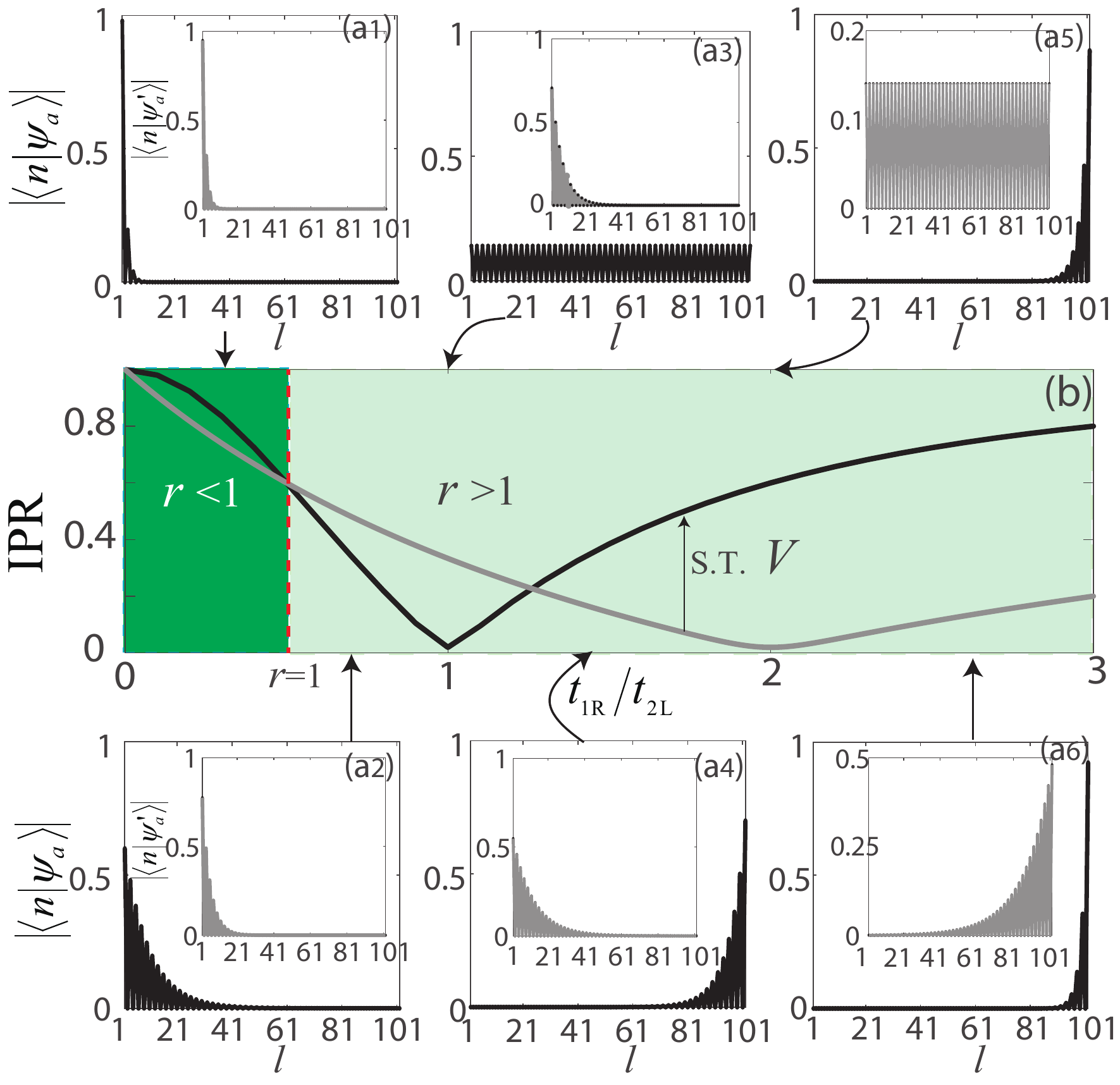}
\caption{  The contrast of zero mode state for the non-Hermitian system $\tilde{\mathcal{H}}$ with that of the Hermitian system $H'$.
Profile of a zero mode of non-Hermitian system $\tilde{\mathcal{H}}$ (main figure) and Hermitian system $H'$ (inset) in subgraph(a1--a6).
(b) The IPR of zero mode  state for the open system terminated with the $A$ site at both ends. The black curve represents the zero mode $|\psi_a\rangle$ of $\tilde{\mathcal{H}}$ and the gray one represents the zero mode $|\psi'_a\rangle$ of $H'$. While the parameters $(t_{1\text{L}},t_{2\text{L}},t_{2\text{R}})=(0.5,0.5,1)$ are fixed, the parameter $t_{1\text{R}}$ takes 0.1 (a1); 0.3 (a2); 0.5 (a3); 0.7 (a4); 1 (a5) and 1.3 (a6), respectively.
 }\label{contrast}
\end{figure}

    For the open system terminated with the $A~(B)$ site at both ends, i.e. the total number of sites is odd, the reflection symmetry of the Hermitian Hamiltonian $H'$ is broken and there is always  a zero-mode state $|\psi'_{0}\rangle=|\psi'_{a(b)}\rangle$ with energy $E=0$, whose wave function only distributes on the $A~(B)$ sublattice. According to analytical analysis, the zero-mode state  $|\psi'_{a}\rangle$  distributes only on the sublattice $A$ and $|\langle n|\psi'_{a}\rangle|$ is proportional to $(\sqrt{t_{1\text{L}}t_{1\text{R}}/t_{2\text{L}}t_{2\text{R}}})^{n-1}$ , which suggests the distribution of  zero mode state would change from left (the inset of Fig.\ref{contrast}(a1--a4)) to right (the inset of Fig.\ref{contrast}(a6))edge when the parameter $t_{1\text{L}}t_{1\text{R}}/t_{2\text{L}}t_{2\text{R}}$ crosses over the the transition point $|t_{1\text{L}}t_{1\text{R}}/t_{2\text{L}}t_{2\text{R}}|=1$ from below. At the transition point, the zero mode wavefuntion would spread over all the lattice (the inset of Fig.\ref{contrast}(a6)). This is verified by the numerical results as shown in Fig.\ref{contrast}(gray), where the IPR of $|\psi'_{a}\rangle$  takes a minimal value at $|t_{1\text{L}}t_{1\text{R}}/t_{2\text{L}}t_{2\text{R}}|=1$ like the gray curve in Fig.\ref{contrast}(b).\\
 \indent By using the similarity transition, the zero-mode state $|\psi_{0}\rangle$ of the non-Hermitian system can be obtained by $|\psi_{a}\rangle =V|\psi'_{a}\rangle$, which distributes only on the sublattice $A$. Accordingly. $|\langle n|\psi_{a}\rangle|$ is proportional to $(r\sqrt{t_{1\text{L}}t_{1\text{R}}/t_{2\text{L}}t_{2\text{R}}})^{n-1}=(t_{1\text{R}}/t_{2\text{L}})^{n-1}$, which suggests the distribution of  zero mode state would change from left (the main figure of Fig.\ref{contrast}(a1) and (a2)) to right (the main figure of Fig.\ref{contrast}(a4)-(a6)) edge when the parameter $t_{1\text{R}}/t_{2\text{L}}$ crosses over the transition point $|t_{1\text{R}}/t_{2\text{L}}|=1$ from below. At the transition point, the zero mode wavefuntion would spread over all the lattice as shown in the main figure of Fig.\ref{contrast}(a3). This is verified by the numerical results as shown in Fig.\ref{contrast}, where the IPR of $|\psi_{a}\rangle$  takes a minimal value at $|t_{1\text{R}}/t_{2\text{L}}|=1$.

{}

%\end{CJK*}

\end{document}